\documentclass[draftclsnofoot,onecolumn]{IEEEtran}
\ifCLASSINFOpdf
\else
\fi
\usepackage{graphicx}
\usepackage{xcolor}
\usepackage{calc}
\usepackage{amssymb}
\usepackage{amsmath}
\usepackage{commath}
\usepackage{citesort}
\usepackage{algorithm}
\usepackage[normalem]{ulem}
\usepackage{algorithmicx,algpseudocode}
\algtext*{EndIf}
\algtext*{EndFor}
\algtext*{EndWhile}

\newcommand\numberthis{\addtocounter{equation}{1}\tag{\theequation}}
\graphicspath{{./} {figures/}}
\usepackage{caption}
\begin{document}
\title{Two-Part Reconstruction with Noisy-Sudocodes}
\author{\IEEEauthorblockN{Yanting Ma,\IEEEauthorrefmark{1}
Dror Baron,\IEEEauthorrefmark{1} and
Deanna Needell\IEEEauthorrefmark{2}}\\
\IEEEauthorblockA{\IEEEauthorrefmark{1}Department of Electrical and Computer Engineering\\
North Carolina State University;
Raleigh, NC 27695, USA\\ 
Email: $\lbrace $yma7, barondror$\rbrace $@ncsu.edu}\\
\IEEEauthorblockA{\IEEEauthorrefmark{2}Department of Mathematical Sciences\\
Claremont McKenna College;
Claremont, CA 91711, USA\\
Email: dneedell@cmc.edu}
\thanks{YM and DB were supported in part by the National Science Foundation under grant CCF-1217749 and in part by the U.S. Army Research Office under grants W911NF-04-D-0003 and W911NF-14-1-0314. DN was supported by the National Science Foundation Career grant $\#1348721$, Simons Foundation
Collaboration grant $\#274305$ and the Alfred P. Sloan Research Fellowship. Portions of the work appeared at the IEEE Global Conference on Signal and Information Processing (GlobalSIP), Austin, TX, Dec. 2013~\cite{MaBaronNeedell2013}.}
}
\IEEEoverridecommandlockouts
\maketitle
\begin{abstract}
We develop a two-part reconstruction framework for signal recovery in compressed sensing (CS), where a fast algorithm is applied to provide partial recovery in Part~1, and a CS algorithm is applied to complete the residual problem in Part~2. Partitioning the reconstruction process into two complementary parts provides a natural trade-off between runtime and reconstruction quality.
To exploit the advantages of the two-part framework, we propose a Noisy-Sudocodes algorithm that performs two-part reconstruction of sparse signals in the presence of measurement noise. 
Specifically, we design a fast algorithm for Part~1 of Noisy-Sudocodes that identifies the zero coefficients of the input signal from its noisy measurements. Many existing CS algorithms could be applied to Part~2, and we investigate approximate message passing (AMP) and binary iterative hard thresholding (BIHT). 
For Noisy-Sudocodes with AMP in Part~2, we provide a theoretical analysis that characterizes the trade-off between runtime and reconstruction quality. 
In a 1-bit CS setting where a new 1-bit quantizer is constructed for Part~1 and BIHT is applied to Part~2, numerical results show that the Noisy-Sudocodes algorithm improves over BIHT in both runtime and reconstruction quality.
\end{abstract}

\section{Introduction}
\label{sec_intro}
In the compressed sensing (CS) signal acquisition paradigm~\cite{DonohoCS,CandesRUP}, 
sparse signals ${\bf x}\in \mathbb{R}^N$ containing only $K\ll N$ nonzero coefficients can be reconstructed from $M<N$ noisy linear measurements of the form ${\bf y} =\Phi {\bf x}+{\bf z}$, where $\Phi\in \mathbb{R}^{M\times N}$, and ${\bf y}$, ${\bf z}\in \mathbb{R}^M$.
While reconstruction quality is an important criterion for algorithm design, the runtime is also of great concern in practical applications.

{\bf Prior art:} There is a vast literature on CS signal reconstruction algorithms~\cite{DonohoCS,C06:Compressive}; many existing algorithms can be classified as combinatorial or geometric.
The combinatorial approach uses sparse and often binary measurement matrices~\cite{Indyk2008,Iwen2014sparsematrice}, and features fast recovery but requires a suboptimal  number of measurements. Sparse binary measurement matrices based on expander graphs have been shown to have good properties for CS reconstruction problems~\cite{jafarpour2009,Raginsky2011expander}. 
The geometric approach often uses dense measurement matrices that satisfy the Restricted Isometry Property (RIP)~\cite{CandesRUP}. Linear programming~\cite{Candes05b} can be applied to perform robust reconstruction with a smaller number of measurements at the expense of greater runtime. Greedy algorithms such as CoSaMP~\cite{Cosamp08} and IHT~\cite{BlumensathDavies2009} offer similar reconstruction quality while requiring less runtime. 

Inference based on message passing was first introduced to CS by Sarvotham et al.~\cite{SudoLDPC,CSBP2010}.
The more recently proposed approximate message passing (AMP) algorithm~\cite{DMM2009} applies the central limit theorem to sum-product belief propagation (BP) (or quadratic approximation to max-sum BP) followed by Taylor expansion to simplify the messages passing between nodes.
AMP has received considerable attention because of its fast convergence as an iterative algorithm and the state evolution (SE)~\cite{DMM2009,Bayati2011} formalism that characterizes the reconstruction problem at each iteration. The theory underlying AMP relies on dense i.i.d. random matrices, which would make the computational complexity of matrix operations in each iteration higher than desired for large signal dimension. Nevertheless, AMP with Fourier and Hadamard matrices~\cite{Javanmard2012} as well as spatially-coupled Fourier and Hadamard matrices~\cite{Barbier2014} have been shown numerically to approximately match the SE derived from i.i.d. random matrices while having lower computational complexity due to special structures in the Fourier and the Hadamard matrices. 

The Sudocodes algorithm~\cite{sudo_isit}, which is related to verification codes~\cite{Luby2005,ZhangPfister2012}, provides an alternative scheme for fast reconstruction of sparse signals when there is no measurement noise. The reconstruction process is partitioned into two parts: Part~1 efficiently recovers most of the zero coefficients and some of the nonzero coefficients from the measurements acquired via a sparse measurement matrix; Part~2 applies a dense measurement matrix and an algorithm with higher computational complexity. Despite the higher complexity, the runtime is still reasonable because Part~2 solves the smaller reconstruction problem left over from Part~1.
A variation of the Sudocodes algorithm is group testing basis pursuit CS (GBCS)~\cite{Talari2011gbcs}, which applies a CS reconstruction algorithm, Basis Pursuit, in Part~2. Sudocodes and GBCS are both fast. However, they can only be applied to noiseless measurements, i.e., ${\bf y}=\Phi{\bf x}$, which is impractical in many real-world applications. Nonetheless, the concept of two-part reconstruction motivates a more practical framework that performs fast reconstruction in the presence of noise.

{\bf Contributions:}
In our earlier work~\cite{MaBaronNeedell2013}, we have generalized the Sudocodes algorithm~\cite{sudo_isit} to a two-part reconstruction framework and proposed a Noisy-Sudocodes algorithm. We designed a zero-identification algorithm for Part~1 of Noisy-Sudocodes, whereas in Part~2 we employed existing CS algorithms. The Noisy-Sudocodes algorithm has been shown to provide promising numerical results in the 1-bit CS framework~\cite{Boufounos2008} where 1 bit is utilized to quantize each entry of the measurements.

In the current work, we present a novel analysis of the Noisy-Sudocodes algorithm. Specifically, we derive a theoretical characterization of the zero-identification algorithm in Part~1, and make the mean square error (MSE) of the entire Noisy-Sudocodes algorithm computable if the algorithm applied to Part~2 can also be theoretically characterized. We select AMP~\cite{DMM2009} as an example of Part~2 to carry out our analysis. This extension of our earlier work allows us to develop a trade-off between the runtime and the reconstruction quality of Noisy-Sudocodes, which highlights the benefit of a two-part framework, and adds practical value to the Noisy-Sudocodes algorithm.

{\bf Organization:}
The remainder of the paper is arranged as follows. We introduce the two-part framework and our proposed Noisy-Sudocodes algorithm in Section \ref{sec.two-part}. A theoretical analysis for Noisy-Sudocodes with AMP in Part~2 is provided in Section~\ref{sec.analysis}. Numerical results for an application of Noisy-Sudocodes to 1-bit CS are presented in Section~\ref{sec.application}, and we conclude the paper in Section~\ref{sec.conclusion}.

\section{Two-Part Reconstruction}
\label{sec.two-part}
\subsection{The two-part framework}
\label{subsec.framework}
In our two-part framework, Part~1 applies a fast algorithm to
recover part of the coefficients of the input signal. The indices of the coefficients that are not recovered in Part~1 are recorded, and Part~2 only processes those remaining coefficients. In the noiseless Sudocodes algorithm~\cite{sudo_isit} where perfect reconstruction is required, the coefficients sent to Part~2 are simply the ones that cannot be perfectly recovered in Part~1. However, when there is measurement noise, a trade-off between runtime and reconstruction quality needs to be considered. On the one hand, we want Part~1 to recover as many coefficients as possible in order to reduce runtime, because Part~2 is in general more complex and slower than Part~1. On the other hand, it is overly ambitious to expect a simple and fast algorithm used in Part 1 to perform high quality
reconstruction, especially in the presence of noise, and so we should not allow Part 1 to reconstruct too many coefficients. 
 
\subsection{Noisy-Sudocodes algorithm}
\label{subsec.algo}
We propose a Noisy-Sudocodes algorithm within the two-part framework. Specifically, we design a fast algorithm to identify the zero coefficients of the input signal, which is suitable for Part~1. The Noisy-Sudocodes algorithm is then defined as a two-part algorithm that applies the zero-identification algorithm to Part~1, and a CS reconstruction algorithm to Part~2. Two examples of CS algorithms that we explore are AMP (Section~\ref{sec.analysis}) and BIHT (Section~\ref{sec.application}).

\begin{algorithm}[t!]
\caption{Sudocodes~\cite{sudo_isit}}
\label{algo:Sudocodes}
\textbf{Inputs:} ${\bf y}_1$, $\Omega^\text{row}$ (support sets of rows of $\Phi_1$), ${\bf y}_2$, $\Phi_2$\\
\textbf{Initialization:} $i=0$, $\widehat{{\bf x}}=0$, $\text{T}=\lbrace 1,2,...,N \rbrace$\\
{\bf Part~1:}
\begin{algorithmic}
\While{$|T|>M_2$ and $i\leq M_1$}
\State $i=i+1$
\If{$y_{1,i}=0$}
\State $\text{T}=\text{T}\setminus\Omega^{\text{row}}_i$
\EndIf
\If{$y_{1,i}\neq 0$}
\For{$k=1:i-1$}
\If{$y_{1,k}=y_{1,i}$}
\State $\text{T}=\text{T}\setminus((\Omega^\text{row}_i\cup\Omega^\text{row}_k)\setminus(\Omega^\text{row}_i\cap\Omega^\text{row}_k))$
\If{$|(\Omega^\text{row}_i\cap\Omega^\text{row}_k)|=1$}
\State $\widehat{x}_{(\Omega^\text{row}_i\cap\Omega^\text{row}_k)}=y_{1,i}$
\State $\text{T}=\text{T}\setminus(\Omega^\text{row}_i\cap\Omega^\text{row}_k)$
\EndIf
\EndIf
\EndFor
\EndIf
\EndWhile
\end{algorithmic}
{\bf Part~2:}\\
$\widetilde{\Phi}=\Phi_{2,\textsf{T}}$\\
$\widehat{{\bf x}}_2=\widetilde{\Phi}^\dagger{\bf y}_2$\\
$\widehat{{\bf x}}_\text{T}=\widehat{{\bf x}}_2$\\
{\bf Outputs:} $\widehat{{\bf x}}$
\end{algorithm}
 
Let ${\bf x}$ be the input signal, and denote the $j$th coefficient of ${\bf x}$ by $x_j$. We assume that ${\bf x}$ is real-valued,\footnote{The extension of our framework to complex-valued signals is left for future work.} and that any subset of the nonzero coefficients of ${\bf x}$ does not sum up to zero with high probability. For most applications of interest, the values of the large components are arbitrary, and the event that several nonzero components sum up to zero is quite unlikely. However, for models in which our assumption fails, our approach would need to be modified.  
Similar to Sudocodes~\cite{sudo_isit}, the measurements of Noisy-Sudocodes are acquired via a sparse measurement matrix $\Phi_1\in\mathbb{R}^{M_1\times N}$ in Part~1 and a dense matrix $\Phi_2\in\mathbb{R}^{M_2\times N}$ in Part~2, so that in total $M=M_1+M_2$ measurements are used.  
Denote the measurement noise in Parts~1~and~2 by ${\bf z}_1$ and ${\bf z}_2$, respectively. The noisy measurement systems in the two parts are given by:
\begin{align}
\text{Part 1: }{\bf y}_1&=\Phi_1 {\bf x}+{\bf z}_1,\label{eq_CSy1}\\
\text{Part 2: }{\bf y}_2&=\Phi_2 {\bf x}+{\bf z}_2.\label{eq_CSy2}
\end{align}
Let $\widehat{{\bf x}}_1$ be the reconstructed signal in Part~1, and denote the $j$th entry of $\widehat{{\bf x}}_1$ by $\widehat{x}_{1,j}$. A set of successive integers $\lbrace 1,...,N\rbrace$ is denoted by $[N]$.  Define $\Omega^\text{row}_i$ and $\Omega^\text{col}_j$ as the support sets (sets of indices of nonzeros) of the $i$th row and the $j$th column of $\Phi_1$, respectively, where $i\in[M_1]$ and $j\in[N]$. Let $\epsilon\geq 0$ be a constant that depends on the noise level.\footnote{We will see how to optimize $\epsilon$ in Section \ref{subsec.tradeoffs}.} Define an index set that contains the indices of small-magnitude measurements as
\begin{equation}
\label{eq.index_small_y}
\Omega^y=\lbrace i:|y_{1,i}|\leq\epsilon, i\in[M_1]\rbrace.
\end{equation} 

The Noisy-Sudocodes algorithm proceeds as follows:

{\bf Part~1:} The measurement vector ${\bf y}_1$ is acquired via (\ref{eq_CSy1}), and thus each $y_{1,i}$ is the summation of a subset of coefficients of ${\bf x}$ that depends on $\Omega^\text{row}_i$. If there were no measurement noise, as in the Sudocodes algorithm~\cite{sudo_isit}, then under our assumptions on the input ${\bf x}$, a zero measurement can only be the summation of zero coefficients.  In other words, if $y_{1,i}=0$, then ${\bf x}_{\Omega^\text{row}_i}= \mathbf{0}$.  However, in the presence of measurement noise, a measurement is (very) unlikely to be precisely zero. Moreover, a small-magnitude measurement could have measured a combination of multiple large-magnitude coefficients, though with small probability $p$. Nevertheless, it is unlikely that a large-magnitude coefficient could appear in multiple small-magnitude measurements (if $p$ is small, then $p^n$ decreases quickly as $n$ increases). The Noisy-Sudocodes algorithm identifies a coefficient to be zero when it is involved in $c$ or more small-magnitude measurements, where $c$ is a tuning parameter that governs the zero-identification criterion.\footnote{We will see how to optimize $c$ in Section \ref{subsec.tradeoffs}.} 
For those coefficients of ${\bf x}$ that do not satisfy the zero-identification criterion, we record their indices in a set $\text{T}$. That is, $\text{T}=\lbrace j: |\Omega^\text{col}_j\cap \Omega^y|<c, j\in [N] \rbrace$, where $|\cdot|$ denotes cardinality.
Unlike Sudocodes~\cite{sudo_isit}, in which some of the nonzero coefficients can be perfectly recovered in Part~1 because the measurements are noiseless, Noisy-Sudocodes leaves the reconstruction of nonzero coefficients for Part~2, where a more robust algorithm is applied.

{\bf Part~2:} Solve the remaining reconstruction problem with a CS algorithm \textsf{F}. The percentage of the zero coefficients that can be identified in Part~1 depends on the noise level and the desired speed-quality trade-off, and so we may still have an underdetermined system in Part~2. In the case where Part 2 is not an underdetermined system, the least squares approach is optimal if the input is deterministic and the measurement noise is Gaussian, whereas the Bayesian approach might be preferable if the input statistics are available or can be learned. The distribution of the measurement matrix $\Phi_2$ depends on the algorithm \textsf{F} applied to Part~2. 
Let ${\bf x}_\text{T}$ represent the coefficients of ${\bf x}$ at the indices $\text{T}$, and $\Phi_{2,\text{T}}$ represent the submatrix formed by selecting columns of $\Phi_2$ at column indices $\text{T}$. The measurement vector ${\bf y}_2$ is acquired via (\ref{eq_CSy2}).  After receiving $\text{T}$ from Part~1, Part~2 first generates $\Phi_{2,\text{T}}$ from $\Phi_2$.
The CS algorithm \textsf{F} then takes $\Phi_{2,\text{T}}$ and ${\bf y}_2$, and computes $\widehat{{\bf x}}_2$, the reconstructed signal of ${\bf x}_\text{T}$.

We complete the reconstruction by assigning $\widehat{{\bf x}}_2$ to the final reconstructed signal $\widehat{{\bf x}}$ at indices $\text{T}$. For completion, we summarize the noiseless Sudocodes algorithm from~\cite{sudo_isit} in Algorithm~\ref{algo:Sudocodes}. Our proposed Noisy-Sudocodes algorithm is summarized in Algorithm \ref{algo:Noisy-Sudocodes}.

\section{Analysis of Noisy-Sudocodes with AMP\\ in Part~2}
\label{sec.analysis}
\subsection{Problem setting}
\label{subsec.setting}
We analyze the Noisy-Sudocodes algorithm in a specific setting. The input signal ${\bf x}\in\mathbb{R}^N$ is i.i.d. sparse Gaussian distributed, $x_j\sim (1-s)\delta(x_j)+s\mathcal{N}(0,1)$, where $s\in(0,1)$ is the sparsity rate, and $\delta(\cdot)$ is the delta function~\cite{Papoulis91}. 

{\bf Part~1:} The sparse measurement matrix  $\Phi_1\in\mathbb{R}^{M_1\times N}$ has i.i.d. Bernoulli entries, $\mathbb{P}(\Phi_{1,ij}\neq 0)=\frac{d}{sN}$, where $d$ is a tuning parameter.\footnote{We will see how to optimize $d$ in Section~\ref{subsec.tradeoffs}.} The measurement noise ${\bf z}_1$ is i.i.d Gaussian distributed, $z_{1,i}\sim\mathcal{N}(0,\sigma_z^2)$.

{\bf Part~2:} The approximate message passing algorithm (AMP) is applied to Part~2. We choose to utilize AMP in our analysis, because the state evolution of AMP~\cite{Bayati2011} provides a convenient tool to accurately characterize the MSE performance of AMP. That said, one can generalize our analysis to other algorithms as well. The measurement matrix $\Phi_2\in\mathbb{R}^{M_2\times N}$ has i.i.d. Gaussian entries, $\Phi_{2,ij}\sim\mathcal{N}(0,1/N)$. The measurement noise ${\bf z}_2$ follows the same distribution as ${\bf z}_1$. 

In order to make the input signal to noise ratio (SNR) in Parts~1 and~2 identical, that is $\|\Phi_1 {\bf x}\|_2^2/\|{\bf z}_1\|_2^2=\|\Phi_2 {\bf x}\|_2^2/\|{\bf z}_2\|_2^2$, the nonzero entries of the Bernoulli matrix $\Phi_1$ are scaled by $\sqrt{\frac{s}{d}}$.

Although we only consider Gaussian noise in our analysis, we believe that Noisy-Sudocodes can be extended to more general noise distributions by applying algorithms that can handle non-Gaussian noise in Part~2. The generalized approximate message passing algorithm (GAMP)~\cite{RanganGAMP2010} is one such algorithm; we leave the extension of Noisy-Sudocodes to other noise distributions for future work.

\begin{algorithm}[t!]
\caption{Noisy-Sudocodes}
\label{algo:Noisy-Sudocodes}
\textbf{Inputs:} ${\bf y}_1$, $\epsilon$, $c$, $\Omega^\text{col}$ (support sets of columns of $\Phi_1$), ${\bf y}_2$, $\Phi_2$\\
\textbf{Initialization:} $\widehat{{\bf x}}=0$, $\text{T}=\emptyset$, $\Omega^y=\emptyset$\\
{\bf Part~1: Apply zero-identification criterion}
\begin{algorithmic}
\For{$i=1:M_1$} 
\If{$|y_{1,i}|<\epsilon$}
\State$\Omega^y=\Omega^y\cup\lbrace i \rbrace$
\EndIf
\EndFor
\For{$j=1:N$}
\If{$|\Omega^\text{col}_j\cap\Omega^y|<c$}
\State$\text{T}=\text{T}\cup\lbrace j \rbrace$
\EndIf
\EndFor
\end{algorithmic}
{\bf Part~2: Apply CS reconstruction algorithm $\textsf{F}$}\\
$\widetilde{\Phi}=\Phi_{2,\textsf{T}}$\\
$\widehat{{\bf x}}_2=\text{F}({\bf y}_2, \widetilde{\Phi})$\\
$\widehat{{\bf x}}_\text{T}=\widehat{{\bf x}}_2$\\
{\bf Outputs:} $\widehat{{\bf x}}$
\end{algorithm}

\subsection{Analysis of Part~1}
\label{subsec.part1}
{\bf Asymptotic independence:} Because only Part~1 will be discussed in this subsection, we drop the subscripts that distinguish Parts~1 and~2. The goal of Part~1 is to identify the zero coefficients of ${\bf x}$. Two types of errors could occur in Part~1. The first is missed detections, which are defined as $\text{MD}=\lbrace  j: x_j=0,\hat{x}_{j}\neq 0,j\in [N] \rbrace$. The second is false alarms, which are defined as $ \text{FA}=\lbrace j: x_j\neq 0,\hat{x}_{j}=0, j\in [N] \rbrace$.
Let $\lbrace I_{ij} \rbrace_{i=1,j=1}^{M,N}=\lbrace I_{ij}: i\in [M],j\in [N] \rbrace$ be a set of binary random variables, where $I_{ij}=1$ if the following two conditions are satisfied: ({\em i}) $|y_i|<\epsilon$ given that the value of the $j$th coefficient is $x_j$ and that the $j$th coefficient is involved in $y_i$; and ({\em ii}) $\Phi_{ij}\neq 0$,  which means that the $j$th coefficient is indeed involved in $y_i$.
Denoting $\mathbb{P}(I_{ij}=1)$ by $P_{\epsilon,d}(x_j)$, we have
\begin{eqnarray}
P_{\epsilon,d}(x_j)&=&\mathbb{P}\left(\abs{y_{i}}<\epsilon,\Phi_{ij}=\sqrt{\frac{s}{d}} \middle|  x_j \right)\nonumber\\
&=&\mathbb{P}\left(|y_i|<\epsilon\middle|\Phi_{ij}=\sqrt{\frac{s}{d}},x_j\right)\mathbb{P}\left(\Phi_{ij}=\sqrt{\frac{s}{d}}\middle| x_j\right)\nonumber\\
&=&\mathbb{P}\left(\left\vert\sum_{k=1}^{N}\Phi_{ik}x_k\right\vert <\epsilon\middle|\Phi_{ij}=\sqrt{\frac{s}{d}},x_j \right)\mathbb{P}\left(\Phi_{ij}=\sqrt{\frac{s}{d}}\right)\nonumber\\
&=&\sum_{n=0}^{N-1}\frac{1}{2}\left(\text{erf}\left( \frac{\epsilon-\sqrt{\frac{s}{d}} x_j}{\sqrt{2\left(\frac{ns}{d}+\sigma_z^2\right)}} \right) -\text{erf}\left( \frac{-\epsilon-\sqrt{\frac{s}{d}} x_j}{\sqrt{2\left(\frac{ns}{d}+\sigma_z^2\right)}} \right) \right)\cdot\nonumber\binom{N-1}{n}\left(\frac{d}{N}\right)^n\left(1-\frac{d}{N}\right)^{N-1-n}\cdot\frac{d}{sN},\label{eq_p_eps_d}
\end{eqnarray}
where $\text{erf}(x)=\frac{2}{\sqrt{\pi}}\int_0^{x}e^{-t^2}\text{d}t$ is the error function. Note that if the nonzero coefficients of ${\bf x}$ do not follow a Gaussian distribution, then the pdf of $y_i$ might not have a simple form. In that case, numerical integration would be needed to compute $P_{\epsilon,d}(x_j)$. We present an example for a sparse Laplace distribution, which is widely utilized as a sparsity promoting prior, in Appendix C.
Further, define the sum of $\lbrace I_{ij} \rbrace_{i=1,j=1}^{M,N}$ along $i$ as
\begin{equation}
S_j=\sum_{i=1}^M I_{ij}, \quad j\in[N].\label{eq.S_j}
\end{equation}

We can now rewrite the zero-identification criterion as $S_j\geq c$, and define the probability of missed detection ($P_{\text{MD}}$) and the probability of false alarm ($P_{\text{FA}}$) as:
\begin{align*}
P_{\text{MD}} &= \mathbb{P}\left(\hat{x}_{j}\neq 0 \mid  x_j = 0\right)=\mathbb{P}\left(S_j<c \mid  x_j=0\right),\\
P_{\text{FA}} &= \mathbb{P}\left(\hat{x}_{j}=0 \mid  x_j\neq 0\right)= \mathbb{P}\left(S_j\geq c\mid x_j\neq 0\right).
\end{align*}

Note that there are subtle dependencies in ${\bf y}$. If a subset of nonzero coefficients of ${\bf x}$ is involved in multiple entries of ${\bf y}$, then the magnitudes of those entries of ${\bf y}$ are not independent. Therefore, for each $j$, $\lbrace I_{ij} \rbrace_{i=1}^{M}$ is not independent along $i$, and thus $S_j$ is a sum of dependent Bernoulli random variables. However, the following Lemma shows that dependencies in ${\bf y}$ vanish under certain conditions.\\
{\bf Lemma~1:} Let the input signal and the measurement matrix of Part~1 be defined in Section~\ref{subsec.setting}, and let $P_{\epsilon,d}(x_j)$ and $S_j$ be defined in~(\ref{eq_p_eps_d})~and~(\ref{eq.S_j}), respectively. In the limit of large systems as the signal dimension $N$ goes to infinity, for each $j\in [N]$, $S_j$ converges to $S_B$ in distribution, where $S_B\sim\text{Binomial}(M,P_{\epsilon,d}(x_j))$.

The proof appears in Appendix A. The main point is that the joint characteristic function of ${\bf y}$ can be factorized as the product of its marginal characteristic functions, which implies that entries of ${\bf y}$ are asymptotically independent, and thus for each $j$ we have that $\lbrace I_{ij} \rbrace_{i=1}^{M}$ is asymptotically independent along $i$. Therefore, $S_j$ converges to a sum of i.i.d. Bernoulli random variables.

Using Lemma~1, $P_{\text{MD}}$ and $P_{\text{FA}}$ can be calculated as follows:
\begin{align*}
P_{\text{MD}}   &=\mathbb{P}\left(\hat{x}_{j}\neq0 \mid  x_j=0\right)\\
         &= \sum_{m=0}^{c-1}\binom{M}{m}P_{\epsilon,d}(0)^m\left(1-P_{\epsilon,d}(0)\right)^{M-m}\numberthis\label{P_MD},\\
P_{\text{FA}} &=\mathbb{P}(\widehat{x}_j=0\mid x_j\neq 0)\\
&=\int_{a\neq 0}\mathbb{P}(\widehat{x}_j=0\mid x_j=a)\mathbb{P}(x_j=a\mid x_j\neq 0)\text{d}a\\   
&=\int_{-\infty}^{\infty} P_{\text{FA}}(a)\frac{1}{\sqrt{2\pi}}e^{-\frac{1}{2}a^2}\text{d}a\numberthis\label{P_FA},\\
P_{\text{FA}}(a)&=\mathbb{P}\left(\hat{x}_{j}=0 \mid  x_j= a\right)\\
         &= 1-\sum_{m=0}^{c-1}\binom{M}{m}P_{\epsilon,d}(a)^m\left(1-P_{\epsilon,d}(a)\right)^{M-m}.
\end{align*}

We can now compute the quantities that might affect the performance of Part~2. The expected length $\widetilde{N}$ and the expected sparsity rate $\widetilde{s}$ of ${\bf x}_\text{T}$ can be calculated as:
\begin{align*}
\widetilde{N}&=N\mathbb{P}\left(\hat{x}_{1,j}\neq 0\right)\\
             &=N\left[ (1-s)P_{\text{MD}} + s(1-P_{\text{FA}}) \right],
\end{align*}
and
\begin{align*}
\widetilde{s} &=\frac{sN(1-P_{\text{FA}})}{\widetilde{N}}\\
          &=\frac{s(1-P_{\text{FA}})}{(1-s)P_{\text{MD}}+s(1-P_{\text{FA}})};
\end{align*}
the distribution of ${\bf x}_\text{T}$, which is denoted by $P_{x_j}\left(a\mid j\in \text{T}\right)$, can be calculated as:
\begin{align*}
P_{x_j}\left(a\mid j\in \text{T}\right) &=\mathbb{P}\left(x_j=a\mid \hat{x}_{1,j}\neq 0\right)\\
                  &=\frac{(1-s)P_{\text{MD}}\delta(a)}{(1-s)P_{\text{MD}}+s(1-P_{\text{FA}})}+\frac{s(1-P_{\text{FA}}(a))\frac{1}{\sqrt{2\pi}}e^{-\frac{1}{2}a^2}}{(1-s)P_{\text{MD}}+s(1-P_{\text{FA}})}\numberthis\label{eq_p_x_T};
\end{align*}
the distribution of ${\bf x}_{\text{FA}}$, which is denoted by $P_{x_j}\left(a\mid j\in \text{FA}\right)$ can be calculated as: 
\begin{align*}
P_{x_j}\left(a\mid j\in \text{FA}\right)&=\mathbb{P}\left(x_j=a\mid \hat{x}_{1,j}= 0,x_j\neq 0\right)\\
                  &=\frac{P_{\text{FA}}(a)\frac{1}{\sqrt{2\pi}}e^{-\frac{1}{2}a^2}}{P_{\text{FA}}};
\end{align*}
and the expected value of the norm of ${\bf x}_{\text{FA}}$ can be calculated as
\begin{align*}
\mathbb{E}[\|{\bf x}_{\text{FA}}\|_2^2] = sN\int_{-\infty}^{\infty} a^2P_{\text{FA}}(a)\frac{1}{\sqrt{2\pi}}e^{-\frac{1}{2}a^2}\text{d}a.\numberthis\label{eq_norm_x_FA}
\end{align*}

\begin{figure}[t!]
\center
\noindent
\includegraphics[width=90mm]{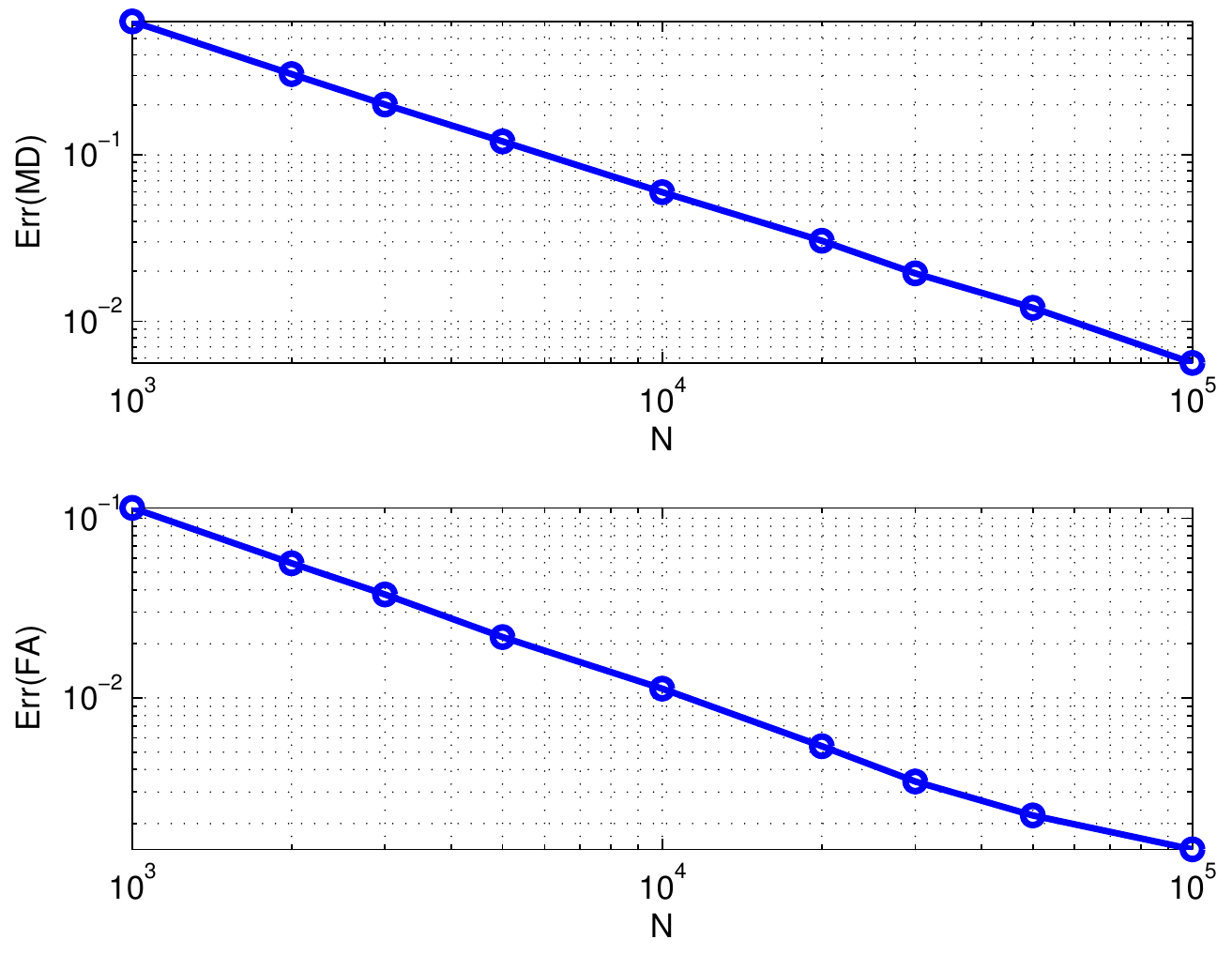}
\caption{Top: Relative error between the empirical and theoretical probability of missed detection. Bottom: Relative error between the empirical and theoretical probability of false alarm. (The theoretical probabilities rely on the asymptotic independence result of Lemma 1.)}
\label{fig.asym_indep01}
\end{figure}

{\bf Numerical verification:} To numerically verify the asymptotic independence property, we simulate Part~1 of Noisy-Sudocodes with different input lengths $N$, and record the empirical probability of missed detection ($P_{\text{MD}}^{\text{em}}$) and the empirical probability of false alarm ($P_{\text{FA}}^{\text{em}}$), 
where we remind the reader that the corresponding theoretical predictions $P_{\text{MD}}$ and $P_{\text{FA}}$ are given by (\ref{P_MD}) and (\ref{P_FA}), and these predictions rely on the asymptotic independence result of Lemma~1. Define the relative error between $P_{\text{MD}}$ and $P_{\text{MD}}^{\text{em}}$ as
\begin{equation*}
\text{Err}(\text{MD}) = \frac{|P_{\text{MD}}-P_{\text{MD}}^{\text{em}}|}{P_{\text{MD}}};
\end{equation*}
the definition of $\text{Err}(\text{FA})$ is similar to that of $\text{Err}(\text{MD})$.
We plot $\text{Err}(\text{MD})$ and $\text{Err}(\text{FA})$ as functions of $N$ in Figure~\ref{fig.asym_indep01}. It is shown in Figure~\ref{fig.asym_indep01} that the error due to the independence assumption in the measurements vanishes at a rate polynomial in $N$. We also obtained similar results for sparse Laplace inputs based on the equations in Appendix C. For brevity, plots are not included.

\subsection{Noisy-Sudocodes with AMP in Part~2}
\label{subsec.part2}
{\bf Gaussianity of noise:} Recall that Part~2 only considers the residual problem left over from Part~1. That is, Part~2 only solves for ${\bf x}$ at the indices $\text{T}$. The missed detection errors in Part~1 result in the zero entries of ${\bf x}_\text{T}$, whereas the false alarm errors in Part~1 result in an extra noise term for Part~2. The extra noise term is generated by ${\bf z}_{\text{FA}}=\Phi_{2,\text{FA}}{\bf x}_{\text{FA}}$, where $\Phi_{2,\text{FA}}$ represents the submatrix formed by selecting columns of $\Phi_2$ at the indices FA. The problem for Part~2 is modeled as
\begin{equation}
{\bf y}_2=\Phi_{2,\text{T}}{\bf x}_\text{T}+{\bf z}_{\text{FA}}+{\bf z}_2\label{eq_Part2_init}.
\end{equation}
Because ${\bf z}_{\text{FA}}$ is a linear mixing of ${\bf x}_{\text{FA}}$, entries of ${\bf z}_{\text{FA}}$ are not independent. However, the following lemma shows that ${\bf z}_{\text{FA}}$ converges to an i.i.d. Gaussian random vector.\\
{\bf Lemma~2:} Let ${\bf z}_{\text{FA}}$ be defined in (\ref{eq_Part2_init}), and $\mathbb{E}[\|{\bf x}_{\text{FA}}\|_2^2]$ be calculated in (\ref{eq_norm_x_FA}). The extra noise term ${\bf z}_{\text{FA}}$ converges to $\bf w$ in distribution, where ${\bf w}\sim\mathcal{N}(0,\sigma_{\text{FA}}^2I)$, and $\sigma_{\text{FA}}^2=\mathbb{E}[\|{\bf x}_{\text{FA}}\|_2^2]/N$.

The proof appears in Appendix B. The main point is that ${\bf z}_{\text{FA}}$ is a sum of i.i.d. random vectors, which converges to a multivariate Gaussian random vector in distribution. It can be shown that ${\bf z}_{\text{FA}}$ has uncorrelated entries. Therefore, ${\bf z}_{\text{FA}}$ converges to an uncorrelated Gaussian random vector.

\begin{figure}[t!]
\center
\noindent
\includegraphics[width=90mm]{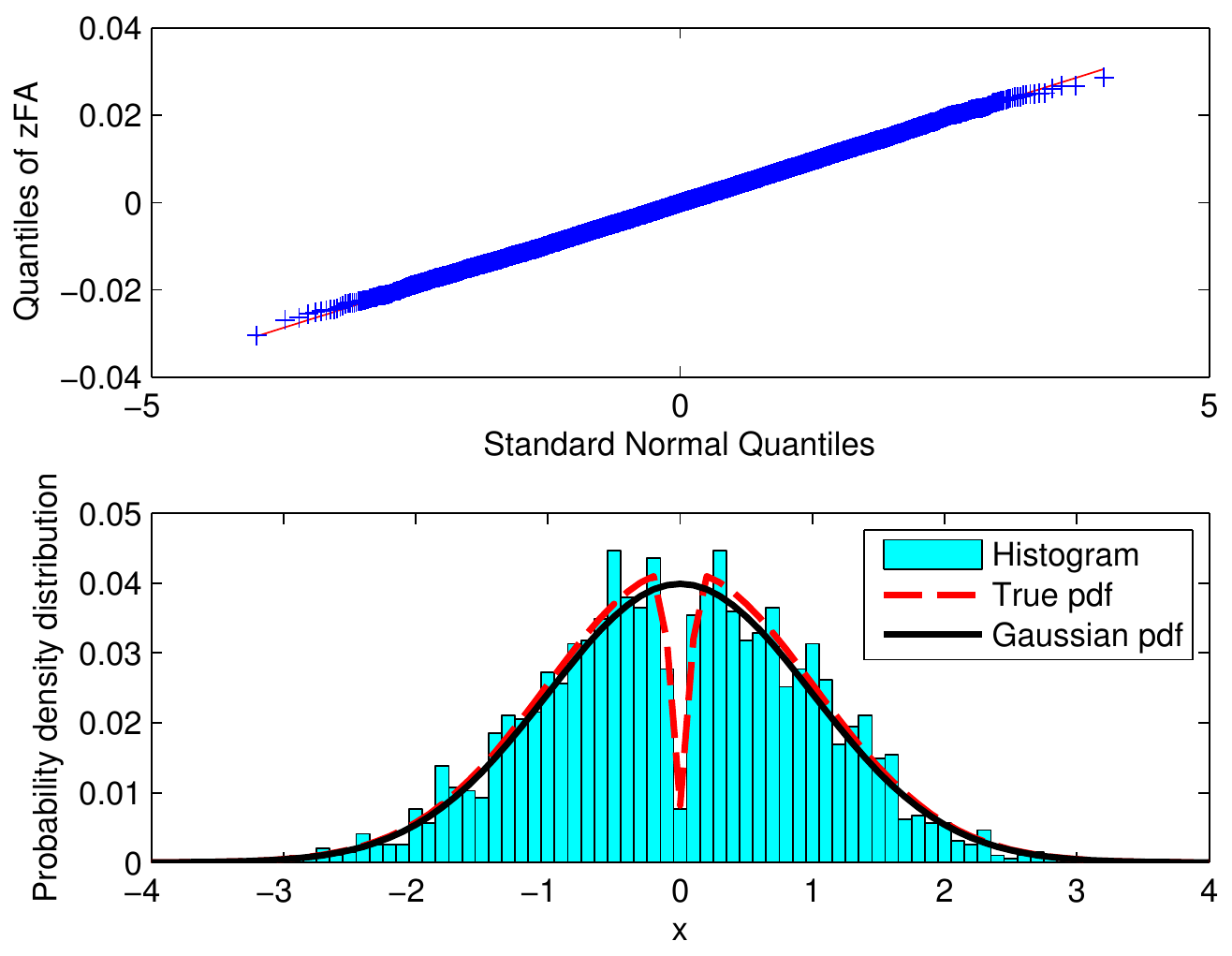}
\caption{Top: QQ plot of the extra noise term ${\bf z}_{\text{FA}}$ due to false alarm errors in Part~1. Bottom: Comparison between the probability density function (pdf) of the nonzero coefficients of the input signal $\widetilde{{\bf x}}$ in Part~2 and the pdf of a standard normal distribution.}
\label{fig.QQplot}
\end{figure}

To numerically verify the Gaussianity of ${\bf z}_{\text{FA}}$, we plot the sample quantiles of ${\bf z}_{\text{FA}}$ versus theoretical quantiles from a normal distribution (QQ plot).
It is shown in the top panel of Figure \ref{fig.QQplot} that the entries of ${\bf z}_{\text{FA}}$ lie on a straight line in the QQ plot, which implies that ${\bf z}_{\text{FA}}$ is marginally Gaussian. 
Next, we test the empirical correlation among the entries of ${\bf z}_{\text{FA}}$, and the resulting empirical correlation is 0.025, which is close to the empirical correlation of an i.i.d. Gaussian random vector of the same length. Therefore, it is verified that ${\bf z}_{\text{FA}}$ converges to an i.i.d. Gaussian random vector.

{\bf Performance analysis with AMP in Part~2:} For notational simplicity, define $\widetilde{{\bf y}}= {\bf y}_2$, $\widetilde{\Phi}\widetilde{{\bf x}}= \Phi_{2,\text{T}}{\bf x}_\text{T}$, and $\widetilde{{\bf z}}= {\bf z}_{\text{FA}}+{\bf z}_2$. Problem (\ref{eq_Part2_init}) can now be rewritten as
\begin{equation}
\widetilde{{\bf y}}=\widetilde{\Phi}\widetilde{{\bf x}}+\widetilde{{\bf z}}\label{eq_Part2_new},
\end{equation}
where $\widetilde{\Phi}\in \mathbb{R}^{M_2\times \widetilde{N}}$ has i.i.d. Gaussian entries, $\widetilde{\Phi}_{ij}\sim \mathcal{N}(0,1/N)$, $\widetilde{{\bf x}}\in \mathbb{R}^{\widetilde{N}}$ is i.i.d. with $\widetilde{x}_j\sim P_{x_j}\left(x\mid j\in \text{T}\right)$ (\ref{eq_p_x_T}), and $\widetilde{{\bf z}}$ is asymptotically i.i.d. Gaussian with zero mean and its variance satisfies $\sigma_{\widetilde{z}}^2=\mathbb{E}[\|{\bf x}_{\text{FA}}\|_2^2]/N+\sigma_z^2$, with $\sigma_z^2$ being the variance of ${\bf z}_2$.

Because $\widetilde{{\bf z}}$ can be approximated as i.i.d. Gaussian noise, we can apply the approximate message passing (AMP) algorithm~\cite{DMM2009} to approximate the minimum mean square error (MMSE) estimate of (\ref{eq_Part2_new}). AMP in Part~2 of Noisy-Sudocodes proceeds as follows:
\begin{align}
{\bf x}^{t+1}&=\eta_t\left(\frac{N}{M_2}\widetilde{ \Phi}^T{\bf r}^t+{\bf x}^t\right),\label{eq.AMP1}\\
{\bf r}^t&={\bf y}-\widetilde{\Phi}{\bf x}^t+\frac{{\bf r}^{t-1}}{\widetilde{R}}\left\langle\eta_{t-1}'\left(\frac{N}{M_2}\widetilde{\Phi}^T{\bf r}^{t-1}+{\bf x}^{t-1} \right)\right\rangle,\label{eq.AMP2}
\end{align}
where $\widetilde{R}=M_2/\widetilde{N}$ is the measurement rate in problem (\ref{eq_Part2_new}), $t$ represents the iteration index, and for a vector ${\bf u}\in\mathbb{R}^N$, $\langle {\bf u} \rangle=\frac{1}{N}\sum_{i=1}^N u_i$. Let ${\bf v}^t=\frac{N}{M_2}\widetilde{\Phi}^T{\bf r}^t+{\bf x}^t$. The scalar estimation function has the form $\eta_t({\bf v}^t)=(\eta_t(v_1^t),...,\eta_t(v_N^t))^T$ as in~\cite{DMM2009}. That is, $\eta_t$ estimates $\widetilde{x}_j$ from its noisy observation $v^t_j$ for each $j\in [\widetilde{N}]$. The derivative of $\eta_t({\bf v}^t)$ is denoted by $\eta_t'({\bf v}^t)$, and $\eta_t'({\bf v}^t)=(\eta_t'(v_1^t),...,\eta_t'(v_N^t))^T$. Due to different measurement matrix normalization schemes, a scaling factor of $N/M_2$ is applied to the AMP updating equations (\ref{eq.AMP1}) and (\ref{eq.AMP2}). It has been rigorously proved~\cite{Bayati2011} that in each iteration, the input of the estimation function $\eta_t$ is equivalent to the noisy observation of $\widetilde{{\bf x}}$ from an additive white Gaussian noise (AWGN) channel. That is, ${\bf v}^t=\widetilde{{\bf x}}+\sigma_t{\bf w}$, where ${\bf w}\sim\mathcal{N}(0,I)$. 
The noise variance $\sigma_t^2$ evolves following the scalar state evolution (SE) formalism~\cite{DMM2009,Bayati2011}:
\begin{equation*}
\sigma^2_{t+1}=\frac{N}{M_2}\sigma^2_{\widetilde{z}}+\frac{1}{\widetilde{R}}\mathbb{E}\left[\left(\eta_t(X+\sigma_tW)-X\right)^2\right],
\end{equation*}
where $X\sim P_{x_j}\left(x\mid j\in \text{T}\right)$ and $W\sim\mathcal{N}(0,1)$.
An unbiased estimator of $\sigma_t^2$~\cite{Montanari2012} can be applied to avoid the calculation of the expected estimation error in each iteration:
\begin{equation*}
\widehat{\sigma}_t^2=\frac{N}{M_2}\frac{\|{\bf r}^t\|_2^2}{M_2}\label{eq.sigma_t}.
\end{equation*}
In order to approximate the MMSE estimate, define the scalar estimation function $\eta_t$ in AMP as the conditional expectation:
\begin{equation}
\eta_t(v_j^t)=\mathbb{E}[\widetilde{x}_j|v_j^t]\numberthis\label{eq_est_func},
\end{equation}
where the prior of $\widetilde{x}_j$ is $P_{x_j}(a|j\in \text{T})$, and the likelihood $P(v^t_j|\widetilde{x}_j)=\mathcal{N}(\widetilde{x}_j,\sigma^2_t)$. Note that when the true distribution of ${\bf x}_\text{T}$~(\ref{eq_p_x_T}) is applied to (\ref{eq_est_func}), the AMP algorithm with i.i.d. random measurement matrix yields the Bayesian optimal reconstruction for (\ref{eq_Part2_new}) in the limit of large systems (i.e., $M_2,\widetilde{N}\rightarrow\infty$ for constant $\widetilde{R}$) for a large region of parameters (signal sparsity, measurement rate, and measurement noise)~\cite{Krzakala2012probabilistic,ZhuBaronCISS2013}.  

We notice that $\widetilde{{\bf x}}$ no longer follows a sparse Gaussian distribution due to the false alarm errors in Part~1. A comparison between the distribution of the nonzero coefficients of $\widetilde{\bf x}$ and a standard normal distribution is shown in the bottom panel of Figure~\ref{fig.QQplot}. Significant discrepancies appear in bins centered around $x=0$, because most false alarm errors occur when the coefficients have small magnitudes. Notice that the entire $\widetilde{\bf x}$ is a sparse signal, which has a probability mass at $x=0$. We might think of $\widetilde{\bf x}$ as a sparse Gaussian signal whose small-magnitude coefficients are approximated as 0, which results in a loss of density around $x=0$ and an increase in the probability mass at $x=0$.
It would be interesting to see how large the performance gap would be if we approximate the prior of $\widetilde{{\bf x}}$ by a sparse Gaussian distribution when calculating the conditional expectation~(\ref{eq_est_func}), because a sparse Gaussian distribution can simplify both the computation and the analysis.

\begin{figure}[t!]
\center
\noindent
\includegraphics[width=90mm]{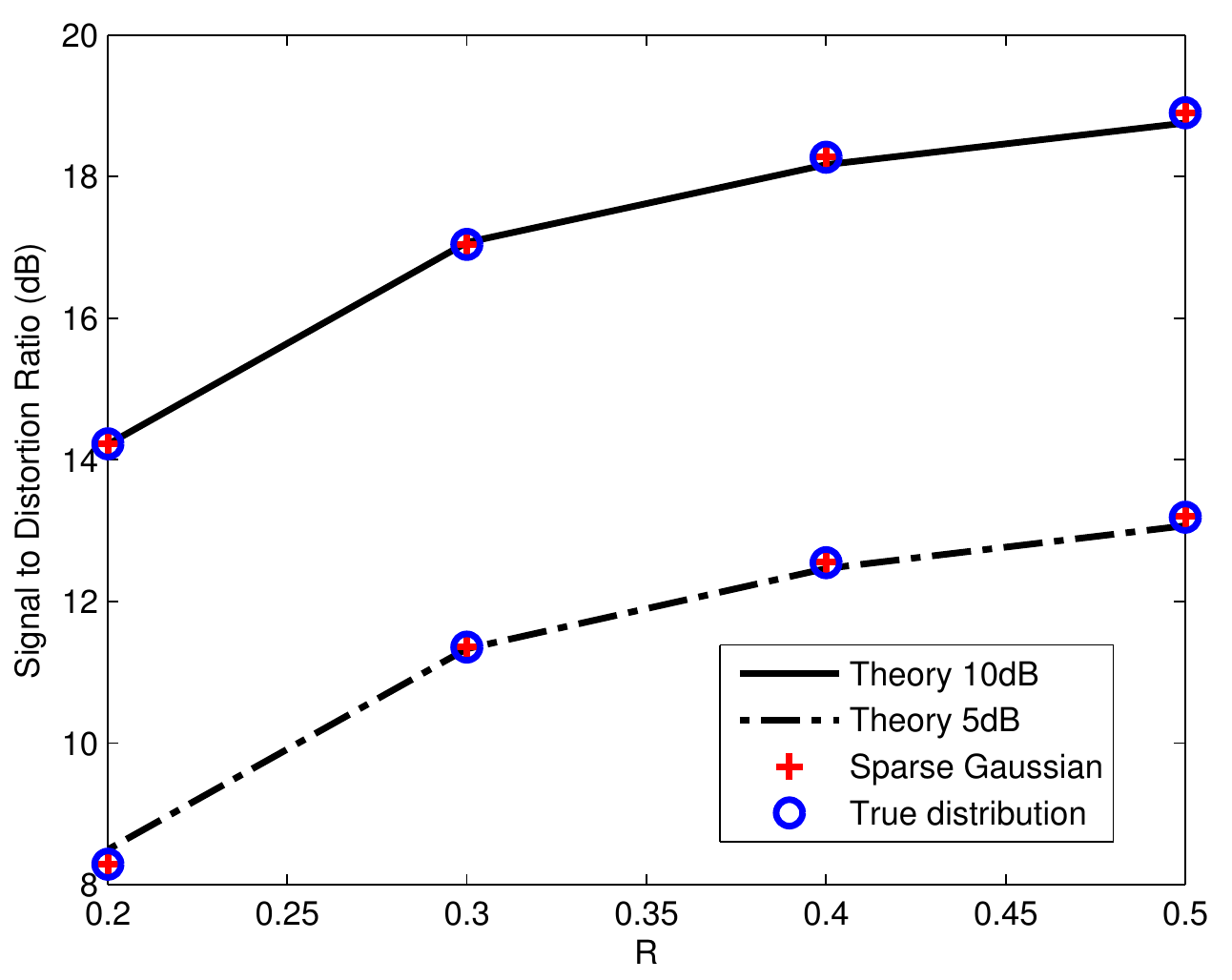}
\caption{Numerical verification of sparse Gaussian approximation to the prior of ${\bf x_\text{T}}$ (\ref{eq_p_x_T}). ($N=20,000$, $s=0.01$, and input SNR $=$ 5 or 10 dB).}
\label{fig.SudoAMP}
\end{figure}

Figure \ref{fig.SudoAMP} compares the signal to distortion ratio (SDR), which is defined as 
\begin{equation}
\label{eq.SDR}
\text{SDR}=10\log_{10}(\mathbb{E}[\|{\bf x}\|_2^2/\|{\bf x}-\widehat{\bf x}\|_2^2]),
\end{equation}
achieved by the theoretical prediction and the numerical results for Noisy-Sudocodes with AMP in Part~2 at different measurement rates $R=M/N=(M_1+M_2)/N$. The prediction for Part~1 follows the analysis in Section \ref{subsec.part1}, and the MMSE for Part~2 (\ref{eq_Part2_new}) applies the replica method for a sparse Gaussian input~\cite{RFG2012,GuoBaronShamai2009}. The empirical results contain: ({\em i}) zero-identification in Part~1 followed by AMP with the sparse Gaussian prior for $\widetilde{{\bf x}}$ in Part~2; ({\em ii}) zero-identification in Part~1 followed by AMP with the true distribution of $\widetilde{{\bf x}}$ in Part~2. 
Figure~\ref{fig.SudoAMP} verifies that it is reasonable to approximate $P_{x_j}(a|j\in \text{T})$~(\ref{eq_p_x_T}) by a sparse Gaussian distribution; any deterioration in reconstruction quality seems minor.
\subsection{Trade-off between runtime and reconstruction quality }
\label{subsec.tradeoffs}
The analysis of the Noisy-Sudocodes algorithm allows us to exploit the advantages provided by its two-part nature. We notice that 4 parameters in the algorithm can be tuned to provide different performances in runtime and reconstruction quality: ({\em i}) the parameter $d$ that governs the sparsity of $\Phi_1$; ({\em ii}) the threshold $\epsilon$ for defining small-magnitude measurements; ({\em iii}) the parameter $c$ that governs the zero-identification criterion; and ({\em iv}) the ratio $r$ of the number of measurements assigned to Part~1 and Part~2. 

It is worth mentioning that the number of AMP iterations could also be tuned. Because AMP is merely one possible example for the algorithm \textsf{F} that can be applied to Part~2, we leave out this tuning parameter in our analysis and fix the number of iterations to be 20, within which AMP generally converges for the numerical settings considered in this paper.

Our goal is to find the parameters $(d,\epsilon,c,r)$ that optimize the trade-off between runtime and reconstruction quality for a given measurement rate.
Both runtime and reconstruction quality are functions of $(d,\epsilon,c,r)$. We have seen how to evaluate the reconstruction quality in terms of SDR (\ref{eq.SDR}) in Sections~\ref{subsec.part1}~and~\ref{subsec.part2}, and let us now model the runtime. Based on the operations performed in the Noisy-Sudocodes algorithm, we model the runtime of Part~1 by
\begin{equation*}
t_1=\alpha_1N+\alpha_2M_1+\alpha_3NM_1,
\end{equation*}
for some ${\boldsymbol \alpha}=(\alpha_1,\alpha_2,\alpha_3)$.
The runtime for Part~2 is modeled as
\begin{equation*}
t_2=\beta_1\widetilde{N}+\beta_2M_2+\beta_3\widetilde{N}M_2,
\end{equation*}
for some ${\boldsymbol \beta}=(\beta_1,\beta_2,\beta_3)$.

We simulate Part~1 with several different values for $N$ and $M_1$, and ${\boldsymbol \alpha}$ is acquired via data fitting with a least squares criterion. We obtain ${\boldsymbol \beta}$ in a similar way. 

The SDR (\ref{eq.SDR}) of Noisy-Sudocodes is evaluated with different parameter values of $(d,\epsilon,c,r)$ at measurement rates $R=M/N\in [0.2,0.9]$. Each set of parameters results in a different $(M_1,M_2,\widetilde{N})$, and thus different $(t_1,t_2)$. The total runtime of Noisy-Sudocodes, $t=t_1+t_2$, is quantized to 30 quantization bins for each $R$, the optimal SDR corresponding to each quantization bin is the highest SDR achieved within that bin, and the parameters that lead to the highest SDR are the optimal parameters. 

\begin{figure}[t!]
\center
\includegraphics[width=90mm]{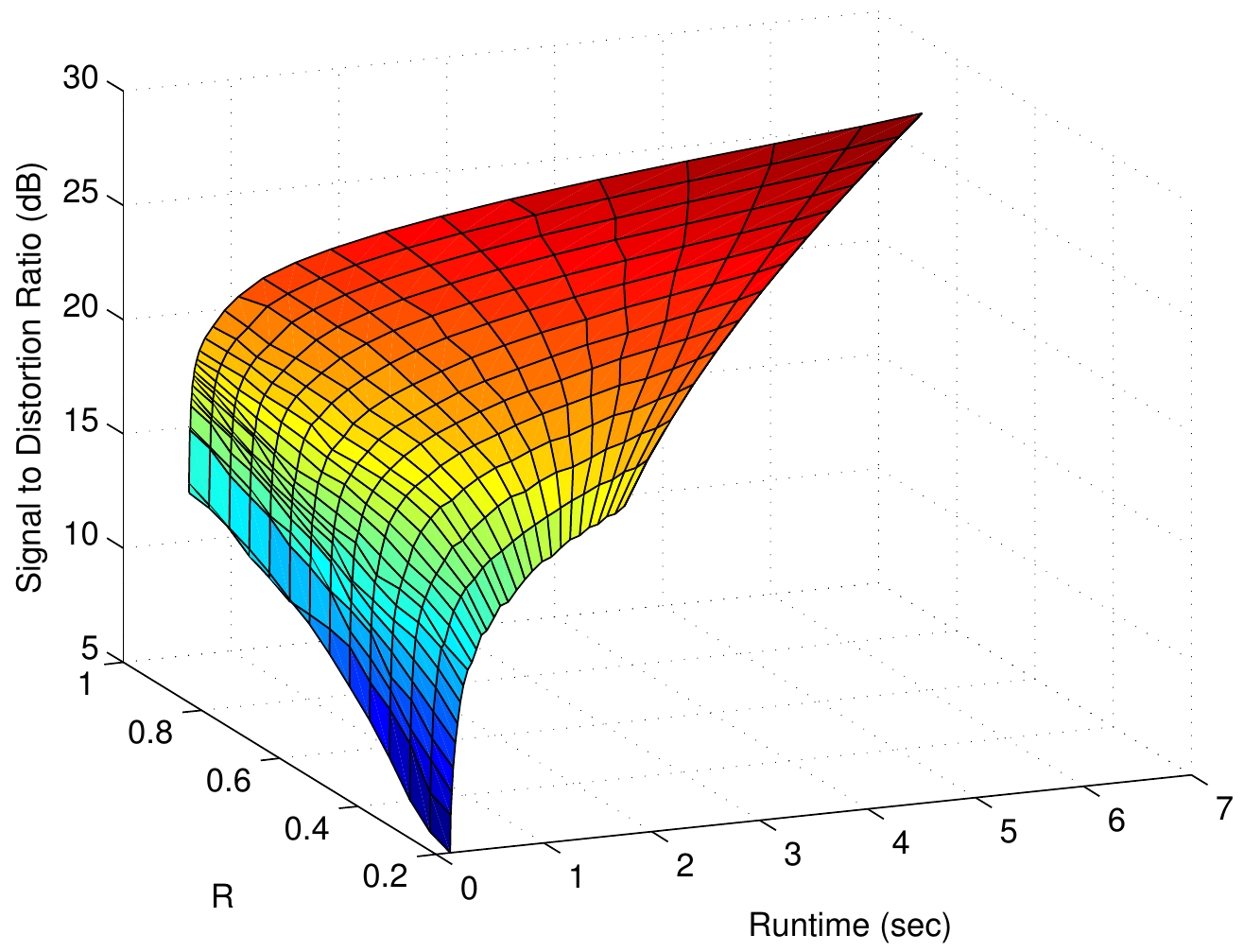}
\caption{Trade-offs between reconstruction quality, measurement rate $R$, and runtime of Noisy-Sudocodes with AMP in Part~2. ($N=20,000$, $s=0.01$, and $\text{input SNR}=10\text{ dB}$).}
\label{fig.tradeoff}
\end{figure}

A plot of SDR as a function of runtime and measurement rate is shown in Figure \ref{fig.tradeoff}. To achieve low runtime, Part~1 needs to be aggressive in identifying zeros, which results in poor reconstruction quality. In the low runtime region, we see a significant improvement in SDR with a small increase in runtime. If we further increase the available runtime, then the high quality algorithm AMP in Part~2 eventually dominates, and thus high SDR is achieved.

\begin{figure}[t!]
\center
\includegraphics[width=90mm]{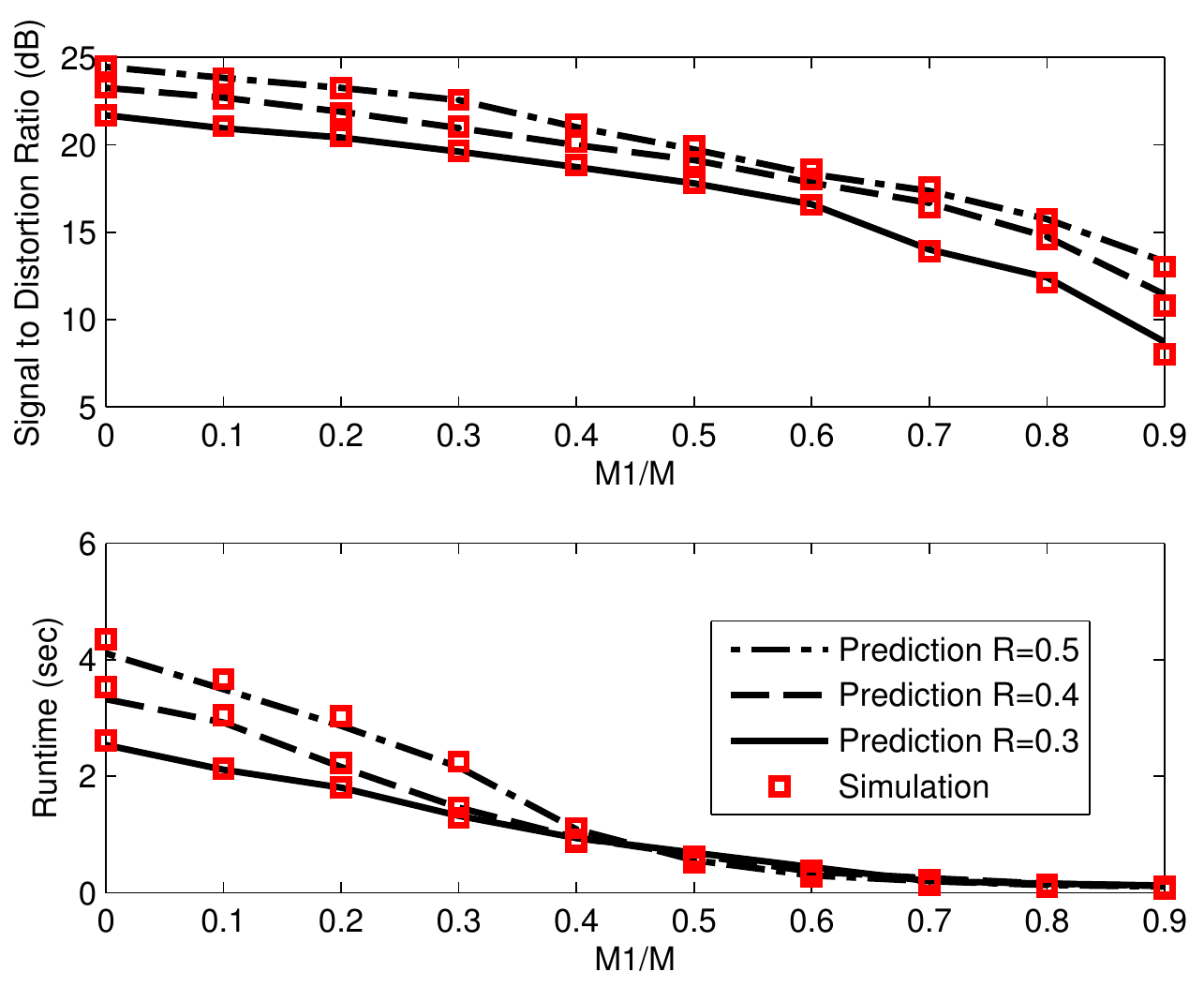}
\caption{Top: Numerical verification of the prediction for SDR~(\ref{eq.SDR}) of Noisy-Sudocodes with AMP in Part~2. Bottom: Numerical verification of the prediction for runtime of Noisy-Sudocodes with AMP in Part~2. ($N=20,000$, $s=0.01$, and $\text{input SNR}=10\text{ dB}$).}
\label{fig.tradeoff_verify}
\end{figure}

To numerically verify the correctness of our predictions of SDR and runtime, we sample some points from Figure~\ref{fig.tradeoff} and set up simulations that utilize the corresponding sets of parameters $(d,\epsilon,c,r)$.
Figure~\ref{fig.tradeoff_verify} shows that our predictions match the simulation results in both SDR and runtime.
\section{Application to 1-bit compressed sensing}
\label{sec.application}
\subsection{Noisy-Sudocodes in 1-bit compressed sensing}
\label{subsec.1bit}
In the previous sections, we discussed Noisy-Sudocodes in CS settings where the measurements are allowed to have infinite quantization resolution. We notice that the fast zero-identification algorithm in Part~1 of Noisy-Sudocodes does not benefit from the high resolution measurements, because we only need to know if the entries of ${\bf y}_1$ are greater or less than $\epsilon$. In other words, the measurements are implicitly quantized to a lower resolution when running Part~1. On the one hand, we see that the fast Part~1 leads to some compromises in reconstruction quality in settings where the measurements are unquantized. On the other hand, Part~1 is not penalized by the loss of quantization resolution in the measurements. This observation naturally leads us to apply Noisy-Sudocodes to a recently proposed 1-bit CS framework~\cite{Boufounos2008}.

In 1-bit CS~\cite{Boufounos2008,Jacques2011robust,PlanVershynin2012,Laska2011trust,Yan2012robust1bit,Yang2013variational}, the measurements are quantized to 1 bit per measurement. The problem model for noiseless and noisy 1-bit CS can be formulated as
\begin{align}
\text{noiseless 1-bit CS:}\quad {\bf y} &= \text{sign} (\Phi {\bf x})\label{eq_noiseless 1 bit},\\
\text{noisy 1-bit CS:}\quad {\bf y} &= \text{sign} (\Phi {\bf x}+{\bf z})\label{eq_noisy 1 bit},
\end{align}
where ${\bf z}$ is the measurement noise before quantization (pre-quantization noise), and
\begin{equation*}
\text{sign}(x)=\begin{cases}
-1, &\text{if } x\leq 0\\
+1, &\text{if } x>0
\end{cases}.
\end{equation*}  

It is interesting to notice that Part~1 of Noisy-Sudocodes motivates a new 1-bit quantizer that performs magnitude quantization.
In particular, we define our proposed 1-bit quantizer as:
\begin{equation}
\label{eq_quantizer}
y_i = \begin{cases}
-1, &\text{if } |(\Phi {\bf x})_i+z_i|\leq\epsilon\\
+1, &\text{if } |(\Phi {\bf x})_i+z_i|>\epsilon
\end{cases}.
\end{equation}
Note that the threshold $\epsilon=0$ when ${\bf z}={\bf 0}$.
If we redefine the index set $\Omega^y$~(\ref{eq.index_small_y}) as
\begin{equation}
\label{eq.index_small_y_1bit}
\Omega^y=\lbrace i\mid |y_{1,i}|=-1, i\in[M_1]\rbrace,
\end{equation}
then Algorithm 1 can be used to solve 1-bit CS reconstruction problems with $\Omega^y$ defined in~(\ref{eq.index_small_y_1bit}) and a 1-bit CS algorithm in Part~2.

A possible 1-bit CS algorithm that can be utilized is binary iterative hard thresholding (BIHT)~\cite{Jacques2011robust}. BIHT often achieves better reconstruction performance than the previous 1-bit CS algorithms in the noiseless 1-bit CS setting. We show by numerical results in Section~\ref{subsec.numerical} that Noisy-Sudocodes with BIHT in Part~2 (Sudo-BIHT) achieves better reconstruction quality than directly applying BIHT. Moreover, Sudo-BIHT is substantially faster than BIHT.

\subsection{Numerical results}
\label{subsec.numerical}
We present simulation results that compare Sudo-BIHT and BIHT in terms of SDR (\ref{eq.SDR}) and runtime in both noiseless and noisy 1-bit CS settings. Runtime is measured in seconds on a Dell OPTIPLEX 9010 running an Intel(R) $\text{Core}^{\text{TM}}$ i7-3770 with 16GB RAM, and the environment is MATLAB R2012a.

The input signal ${\bf x}$ follows a sparse Gaussian distribution with sparsity rate $s=0.005$.
Because the amplitude information of the measurements is lost due to 1-bit quantization, it is usually assumed in the 1-bit CS framework that $\|{\bf x}\|_2^2=1$.
Let $M_1$ and $M_2$ be the number of measurements for Parts~1 and~2 of Sudo-BIHT. Therefore, $M=M_1+M_2$ is the number of measurements for BIHT. The measurement rate $R=M/N$ is set to be within the range $(0,2)$, which is the same range utilized in the paper where BIHT is proposed~\cite{Jacques2011robust}.
Note that in 1-bit CS, we are interested in the number of quantization bits rather than the number of measurements. Therefore, the measurement rate is allowed to be greater than 1. 
In our simulation, we choose $M_1$ such that more than 90 percent of the zero coefficients can be identified in Part~1. The measurement matrix $\Phi_1 \in \mathbb{R}^{M_1\times N}$ is i.i.d. Bernoulli distributed with $\mathbb{P}(\Phi_{1,ij}\neq 0)=\frac{d}{sN}$, where the parameter $d$ is determined numerically. Note that the nonzero entries of the Bernoulli matrix are scaled by $\sqrt{\frac{sN}{d}}$ in order to have the same input SNR as in BIHT. The matrix $\Phi _2 \in \mathbb{R}^{M_2 \times N}$ has i.i.d. Gaussian entries, $\Phi _{2,ij} \sim  \mathcal{N}(0,1)$.  

For BIHT, the measurement matrix $\Phi\in \mathbb{R}^{M\times N}$ has i.i.d. Gaussian entries, $\Phi_{ij}\sim\mathcal{N}(0,1)$.

Finally, the pre-quantization noise ${\bf z}$, which we use in the noisy setting, is i.i.d. Gaussian distributed with zero mean and its variance is $10^{-2.5}.$

{\bf Noiseless setting:} BIHT-$\ell_1$~\cite{Jacques2011robust}, in which the $\ell_1$-norm is utilized in the objective function of the optimization problem solved by BIHT, is applied to the noiseless setting. The measurement vector ${\bf y}_1$ for Part~1 of Sudo-BIHT is acquired via (\ref{eq_quantizer}) with ${\bf z}={\bf 0}$ and $\epsilon=0$, and the measurement vectors ${\bf y}_2$ for Part~2 of Sudo-BIHT and ${\bf y}$ for BIHT  are acquired via (\ref{eq_noiseless 1 bit}). In the noiseless setting, if any entry $y_{1,i}$ only measures zero coefficients, then $y_{1,i}$ will be strictly zero. Therefore, we set $c=1$ in the zero-identification criterion. Note that Part~1 does not introduce any error in the noiseless setting. We iterate over BIHT until the consistency property\footnote{We say that the consistency property of BIHT~\cite{Boufounos2008} is satisfied if applying the measurement and quantization system (\ref{eq_noiseless 1 bit}) and (\ref{eq_noisy 1 bit}) to the reconstructed signal $\widehat{{\bf x}}$ yields the same measurements ${\bf y}$ as the original measurements.} is satisfied or the number of iterations reaches 100.
\begin{figure}[t!]
\center
\noindent
\includegraphics[width=90mm]{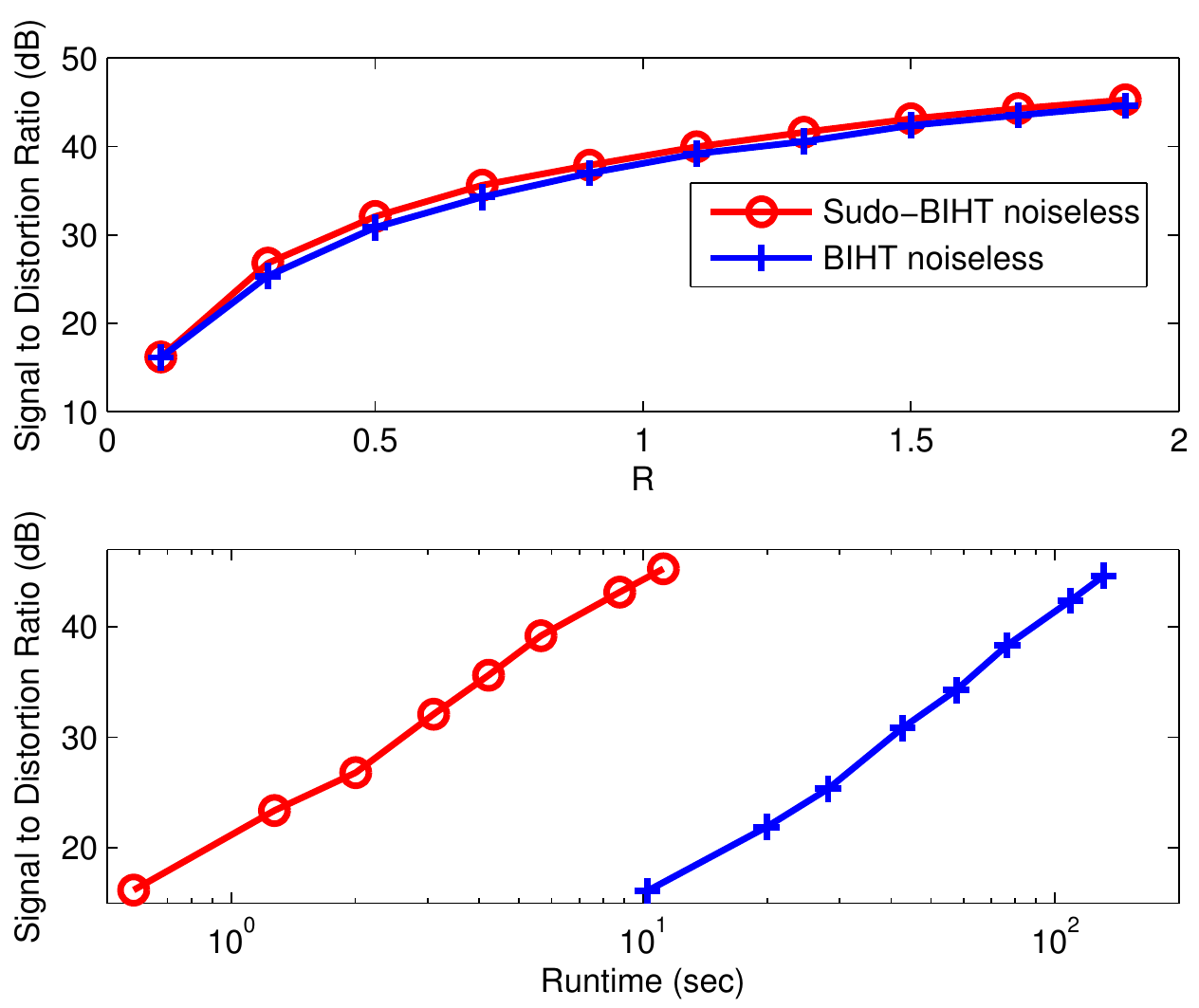}
\caption{Numerical results of Noisy-Sudocodes with BIHT in Part~2 in a noiseless 1-bit CS setting. Top: SDR~(\ref{eq.SDR}) as a function of measurement rate $R$. Bottom: SDR as a function of runtime. ($N=10,000$, $s=0.005$, $c=1$, $d=0.8$, $\epsilon=0$, $M_1/N=0.1$, and $M_2=M-M_1$).}
\label{fig.noiseless}
\end{figure}

In the top panel of Figure~\ref{fig.noiseless}, we plot SDR as a function of the measurement rate $R$. The plot shows that Sudo-BIHT achieves slightly higher SDR than BIHT. As $R$ increases, the SDR for both algorithms increases similarly. Note that the measurements acquired in noiseless 1-bit CS include quantization noise. The quantization noise explains why the SDR achieved in the noiseless 1-bit CS setting is finite, whereas unquantized noiseless measurements yield perfect reconstruction~\cite{DonohoCS,CandesRUP}. In the bottom panel of Figure~\ref{fig.noiseless}, we plot SDR as a function runtime. Note that Sudo-BIHT can achieve the same SDR as BIHT despite running an order of magnitude faster.

{\bf Noisy setting:} BIHT-$\ell_2$~\cite{Jacques2011robust}, in which the $\ell_2$-norm is utilized in the objective function, is applied to the noisy setting. Note that BIHT-$\ell_2$ is more robust to pre-quantization noise than BIHT-$\ell_1$. The measurement vector ${\bf y}_1$ for Part~1 of Sudo-BIHT is acquired via (\ref{eq_quantizer}) with $\epsilon > 0$, and the measurement vectors ${\bf y}_2$ for Part~2 of Sudo-BIHT and ${\bf y}$ for BIHT are acquired via~(\ref{eq_noisy 1 bit}). We set $c=3$, $d=0.8$, and $\epsilon=0.08$ in our simulations because they lead to sufficiently good performance in the sense that Sudo-BIHT improves over BIHT in both runtime and reconstruction quality.

\begin{figure}[t!]
\center
\noindent
  \includegraphics[width=90mm]{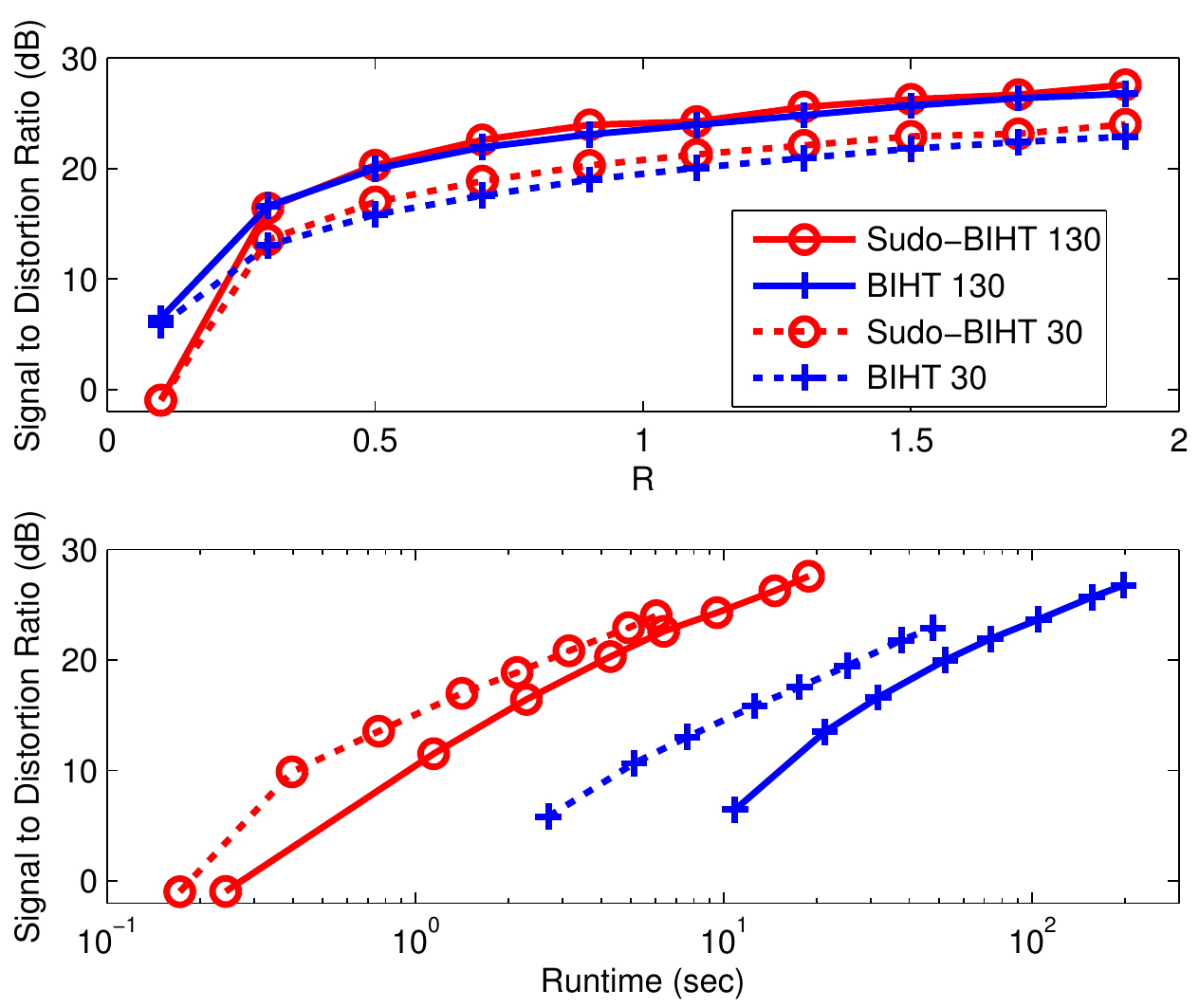}
  \caption{Numerical results of Noisy-Sudocodes with BIHT in Part~2 in a noisy 1-bit CS setting. Top: SDR~(\ref{eq.SDR}) as a function of measurement rate $R$. Bottom: SDR as a function of runtime.  ($N=10,000$, $s=0.005$, $c=3$, $d=0.8$, $\epsilon=0.08$, $M_1/N=0.1$, and $M_2=M-M_1$).}
\label{fig.noisy}
\end{figure}

The resulting SDR versus measurement rate $R$ is shown in the top panel of Figure~\ref{fig.noisy}. 
When the number of iterations for BIHT is 30 in both Part~2 of Sudo-BIHT and BIHT, Sudo-BIHT yields better consistency and thus provides better reconstruction quality. With more iterations, the SDR for both Sudo-BIHT and BIHT improves. The SDR curve of BIHT tends to get closer to Sudo-BIHT as the number of iterations increases, because for Sudo-BIHT, the error introduced in Part 1 cannot be corrected by Part 2. 
We notice that Sudo-BIHT with 130 BIHT iterations (red solid line with circles) improves over BIHT with 30 iterations (blue dotted line with crosses) by roughly 5 dB for the same measurement rate, and the bottom panel of Figure~\ref{fig.noisy} shows that the red solid line with circles can be 5 dB above the blue dotted line with crosses despite requiring approximately half of the runtime.
In other words, problem size reduction due to zero-identification in Part~1 allows BIHT in Part~2 to run more iterations to improve reconstruction quality with reasonable runtime.
\section{conclusion}
\label{sec.conclusion}
We introduced a two-part reconstruction framework that partitions the reconstruction process into two complementary parts. The partitioning leads to a trade-off between runtime and reconstruction quality. Applications such as real-time signal processing where speed is crucial, whereas quality is less important, might benefit from our algorithm. For example, in real-time audio or video processing, delay in time might be more undesirable than deterioration in reconstruction quality.
A Noisy-Sudocodes algorithm was proposed within the two-part framework. Part~1 of Noisy-sudocodes is the zero-identification algorithm, whereas various CS reconstruction algorithms can serve as candidates for Part~2. 
We analyzed the speed-quality trade-off of Noisy-Sudocodes with AMP~\cite{DMM2009} in Part~2 based on the theoretical characterization that we derived for Part~1 and the well-established asymptotic properties of AMP. 
Moreover, numerical results for Noisy-Sudocodes with our 1-bit magnitude-quantizer in Part~1 and BIHT~\cite{Jacques2011robust} in Part~2 imply that Noisy-Sudocodes could be promising for algorithm design in 1-bit CS reconstruction problems.
 
\section*{Appendix A: proof of Lemma~1}
We will show that with the problem setting described in Section~\ref{subsec.setting}, ${\bf y}_1$ is asymptotically independent in the limit of large $N$. The subscript that represents Part~1 is dropped in the following analysis. Denote the characteristic function of ${\bf x}$ by $\Psi_x(t)=\mathbb{E}\left[ e^{itx} \right]$. We will show that for any constant $m\leq M$,
\begin{equation}
\lim_{N\rightarrow\infty} \Psi_{y_1...y_m}(t_1,...,t_m)=\lim_{N\rightarrow\infty} \Psi_{y_1}(t_1)...\Psi_{y_m}(t_m),\label{eq.charFunc}
\end{equation}
where $\Psi_{y_1...y_m}(t_1,...,t_m)= \mathbb{E}\left[ e^{it_1y_1+...+it_cy_m} \right]$ is the joint characteristic function, and expectation is taken with respect to the joint probability density $P(y_1,...,y_m)$. The joint characteristic function can be factorized as the product of the marginal characteristic functions as described in (\ref{eq.charFunc}) if and only if $y_1,...,y_m$ are independent~\cite{Applebaum2005}.

To lighten the notation, we assume that the nonzero entries of the Bernoulli matrix $\Phi$ are ones (we adjusted the nonzero entries in the body of the paper to make the input SNR in Parst~1~and~2 identical), and we ignore the i.i.d. measurement noise.\footnote{Note that if entries of ${\bf y}=\Phi {\bf x}$ are independent, then after adding an i.i.d. noise vector ${\bf z}$, entries of ${\bf y}'={\bf y}+{\bf z}$ are still independent. Therefore, these simplifications do not affect the independence relation among entries of ${\bf y}$.} Under these simplifications, the signal model is
\[
{\bf y}=\Phi {\bf x}=\Phi_{*1}x_1+\Phi_{*2}x_2+...+\Phi_{*N}x_N,
\]
where $\Phi_{*j}$ represents the $j$th column of $\Phi$. 
Define a sequence of random vectors $v_j= \Phi_{*j}x_j$, $j\in [N]$. Notice that $\lbrace v_j \rbrace_{j=1}^{N}$ are i.i.d. random vectors, and thus the characteristic function of the first $m$ entries of ${\bf y}$ is
\begin{equation*}
\Psi_{y}(t_1,...,t_m)=\left( \Psi_{v_1}(t_1,...,t_m)\right)^N.
\end{equation*}
It can be calculated that the characteristic function of a Gaussian random variable with probability density function (pdf) $\mathcal{N}(\mu,\sigma^2)$ is $e^{i\mu t-\frac{1}{2}\sigma^2 t^2}$. 

To establish (\ref{eq.charFunc}), it suffices to show that 
\begin{equation}
\lim_{N\rightarrow\infty} \Psi_{y_1}(t_1)...\Psi_{y_m}(t_m)=e^{d\left( e^{-\frac{1}{2}t_1^2}+...+e^{-\frac{1}{2}t_m^2}-m \right)}\label{eq.charFunc_marginal}
\end{equation}
and
\begin{equation}
\lim_{N\rightarrow\infty} \Psi_{y_1...y_m}(t_1,...,t_m)=e^{d\left( e^{-\frac{1}{2}t_1^2}+...+e^{-\frac{1}{2}t_m^2}-m \right)}\label{eq.charFunc_joint}.
\end{equation}

First, we show (\ref{eq.charFunc_marginal}).
For $m=1$, $v_1$ is a scalar. Recall that the Bernoulli parameter of the Bernoulli matrix in Part~1 is $\frac{d}{sN}$ and the sparsity rate of ${\bf x}$ is $s$. Let $g(x)$ denote the pdf of a Gaussian random variable $x$ with mean 0 and variance 1. Denoting the probability distribution of $v_1$ by $P_{v_1}(u_1)$, we have
\begin{align*}
P_{v_1}(u_1)    &= \frac{d}{N}g(u_1) + \left( 1-\frac{d}{N}\right)\delta(u_1),\\
\Psi_{v_1}(t_1) &= 1-\frac{d}{N} + \frac{d}{N}e^{-\frac{1}{2}t_1^2},\\
\Psi_{y_1}(t_1)   &= \left( 1-\frac{d}{N}+\frac{d}{N}e^{-\frac{1}{2}t_1^2} \right)^N,\\                
\lim_{N\rightarrow \infty} \Psi_{y_1}(t_1) &=\lim_{N\rightarrow \infty} \left( 1+\frac{d}{N}\left( e^{-\frac{1}{2}t_1^2}-1\right) \right)^N\\
&=e^{ d\left( e^{-\frac{1}{2}t_1^2}-1 \right)}.
\end{align*}
Because $\displaystyle \lim_{\substack{N\rightarrow\infty}}\Psi_{y_i}(t_i)$ exists for every $i\in [M]$, for any finite constant $m$, we have
\begin{align*}
\lim_{N\rightarrow\infty}\Psi_{y_1}(t_1)...\Psi_{y_{m}}(t_{m}) &=\lim_{N\rightarrow\infty}\Psi_{y_1}(t_1)...\lim_{N\rightarrow\infty}\Psi_{y_{m}}(t_{m})\\
&=e^{ d\left( e^{-\frac{1}{2}t_1^2}+...+e^{-\frac{1}{2}t_{m}^2}-m\right)}.
\end{align*}
Therefore, (\ref{eq.charFunc_marginal}) is verified.

Second, we show (\ref{eq.charFunc_joint}).
For $m=2$, $v_1$ is a vector of length 2. Denoting the probability distribution of $v_1$ by $P_{v_1}(u_1,u_2)$, we have
\begin{eqnarray*}
P_{v_1}(u_1,u_2)&=&P_{v_1}\left(u_1,u_2\mid x_1=0\right)\mathbb{P}(x_1=0)+P_{v_1}\left(u_1,u_2\mid x_1\neq 0\right)\mathbb{P}\left(x_1\neq 0\right)\\
&=&P_{v_1}\left(u_1,u_2\mid x_1=0\right)\mathbb{P}(x_1=0)+[P_{v_1}(u_1,u_2|\Phi_{11}=0,\Phi_{12}=0,x_1\neq 0)\mathbb{P}(\Phi_{11}=0,\Phi_{12}= 0)\\
&+&P_{v_1}(u_1,u_2|\Phi_{11}=0,\Phi_{12}\neq 0,x_1\neq 0)\mathbb{P}(\Phi_{11}=0,\Phi_{12}\neq 0)\\
&+&P_{v_1}(u_1,u_2|\Phi_{11}\neq 0,\Phi_{12}=0,x_1\neq 0)\mathbb{P}(\Phi_{11}\neq 0,\Phi_{12}=0)\\
&+&P_{v_1}(u_1,u_2|\Phi_{11}\neq 0,\Phi_{12}\neq 0,x_1\neq 0)\mathbb{P}(\Phi_{11}\neq 0,\Phi_{12}\neq 0)]\mathbb{P}(x_1\neq 0)\\
&=&\delta(u_1)\delta(u_2)\left(1-s\right)+\Bigg[ \left(1-\frac{d}{sN}\right)^2\delta(u_1)\delta(u_2)+\frac{d}{sN}\left(1-\frac{d}{sN}\right)\Bigg(\delta(u_1)g(u_2)+\delta(u_2)g(u_1)\Bigg)\\
&+&\left(\frac{d}{sN}\right)^2\delta(u_2-u_1)g(u_1)\Bigg]s \\
&=&\left( 1+\frac{d^2}{sN^2}-\frac{2d}{N} \right)\delta(u_1)\delta(u_2)+\left( \frac{d}{N}-\frac{d^2}{sN^2}\right)\Bigg( \delta(u_1)g(u_2)+\delta(u_2)g(u_1) \Bigg)+\frac{d^2}{sN^2}\delta(u_2-u_1)g(u_1),
\end{eqnarray*}
\begin{align*}
\Psi_{v_1}(t_1,t_2)&=\mathbb{E}\left[ e^{it_1u_1+it_2u_2}\right]\\
                    &=\int_{-\infty}^{\infty}\int_{-\infty}^{\infty} e^{it_1u_1+it_2u_2} P_{v_1}(u_1,u_2) \text{d}u_1\text{d}u_2\\
                    &=\left( 1+\frac{d^2}{sN^2}-\frac{2d}{N} \right) + \left( \frac{d}{N}-\frac{d^2}{sN^2}\right)\left( e^{-\frac{1}{2}t_1^2} + e^{-\frac{1}{2}t_2^2} \right) + \frac{d^2}{sN^2}e^{-\frac{1}{2}(t_1+t_2)^2}\\
                    &=1+\frac{d}{N}\left( e^{-\frac{1}{2}t_1^2}+e^{-\frac{1}{2}t_2^2}-2\right)+\frac{d^2}{sN^2}\left(e^{-\frac{1}{2}(t_1+t_2)^2}-e^{-\frac{1}{2}t_1^2}-e^{-\frac{1}{2}t_2^2}+1 \right),
\end{align*}
\begin{align*}
\lim_{N\rightarrow \infty}  \Psi_{y_1y_2}(t_1,t_2) &= \lim_{N\rightarrow \infty}  \left( \Psi_{v_1}(t_1,t_2) \right)^N\\
&=e^{ d\left( e^{-\frac{1}{2}t_1^2} + e^{-\frac{1}{2}t_2^2} -2 \right) }.
\end{align*}
Similarly, it can be shown for any $m\leq M$ that
\begin{align*}
\lim_{N\rightarrow\infty}\Psi_{y_1...y_m}(t_1,...,t_m)&=\lim_{N\rightarrow\infty}\left( \Psi_{v_1}(t_1,...,t_m) \right)^N\\
             &=e^{ d\left( \sum_{i=1}^ce^{-\frac{1}{2}t_i^2}-m \right)}.   
\end{align*}
Therefore, (\ref{eq.charFunc_joint}) is also verified, which establishes (\ref{eq.charFunc}).

We conclude that in the limit of large $N$, for each $j\in[N]$, the indicator variables $\lbrace I_{ij}\rbrace_{i=1,j=1}^{M,N}$ are independent along $i$.
Therefore, $S_j=\sum_{i=1}^{M}I_{ij}$ converges to a Binomial random variable $S_B$ in distribution~\cite{Klenke2013}, where $S_B\sim\text{Binomial}\left(M,P_{\epsilon,d}(x_j)\right)$. 
\section*{Appendix B: proof of Lemma~2}
To simplify the notation, we drop the subscripts of $\Phi_{2,\text{FA}}$ and $x_{\text{FA}}$, and let $\Phi$ represent the submatrix formed by columns of $\Phi_2$ at the indices $\text{FA}$, and ${\bf x}$ represent entries of ${\bf x}$ at the indices $\text{FA}$, where FA represents the false alarms defined in Section~\ref{subsec.part1}.
Define a sequence of vectors $v_j=\Phi_{*j}x_{j}$, $j\in[|\text{FA}|]$.

We notice that ${\bf z}_{\text{FA}}$ is a sum of i.i.d. random vectors, and the components in each vector are uncorrelated. That is, 
\begin{align*}
{\bf z}_{\text{FA}} &=\sum_{j=1}^{|\text{FA}|} v_j,\\
\mathbb{E}[v_{jt}v_{js}]&=\mathbb{E}[(x_{j}\Phi_{jt})(x_{j}\Phi_{js})]\\
               &=\mathbb{E}[x_{j}^2]\mathbb{E}[\Phi_{jt}\Phi_{js}]\\
               &=\begin{cases}
               0,\quad &\text{if } s\neq t\\
               \mathbb{E}[x_{\text{FA},j}^2]/N,\quad &\text{if } s=t 
               \end{cases} .     
\end{align*}
The proof is completed by applying the Multivariate Central Limit Theorem.\\
{\bf Central Limit Theorem in $\mathbb{R}^d$~\cite{Klenke2013}:} Let $(X_n)_{n\in\mathbb{N}}$ be i.i.d. random vectors with $\mathbb{E}[X_{n,i}]=0$ and $\mathbb{E}[X_{n,i}X_{n,j}]=C_{ij}$, $i,j\in[d]$. Let $S^*_n=\frac{X_1+...+X_n}{\sqrt{n}}$. Then
\[P_{S^*_n}\rightarrow \mathcal{N}(0,C)\quad \text{in distribution}.\]  

By the multivariate central limit theorem, the distribution of the vector ${\bf z}_{\text{FA}}$ converges to $\mathcal{N}(0,C_{vv})$, where $C_{vv}$ is a diagonal covariance matrix with $\mathbb{E}[\|{\bf x}_{\text{FA}}\|_2^2]/N$ on its diagonal. Therefore, ${\bf z}_{\text{FA}}$ converges to an i.i.d. Gaussian random vector in distribution.

\section*{Appendix C: Equations for sparse Laplace input}

Let $y=\sum_{j=1}^n\sqrt{\frac{s}{d}} x_j + z$, where $x_1,x_2,...,x_n$ are i.i.d. standard Laplace random variables with pdf 
$f(x)=\frac{1}{2b}\exp\left(-\frac{|x-\mu|}{b}\right)$, where $\mu=0$, $b=1$, and $z\sim\mathcal{N}(0,\sigma^2)$. It can be calculated that the characteristic function with respect to $f(x)$ is $\Psi_x(t)=\frac{e^{ixt}}{(1+b^2t^2)}$.

We now have that
\begin{align*}
\Psi_{\sqrt{\frac{s}{d}}x}(t) &= \frac{1}{1+\frac{s}{d} t^2},\\
\Psi_z(t) &= e^{-\frac{1}{2}\sigma^2t^2},\\
\Phi_y(t) &=\Psi_x(t)^n\Psi_z(t)\\
          &=\left(\frac{1}{1+\frac{s}{d} t^2}\right)^ne^{-\frac{1}{2}\sigma^2t^2},\\
f(y)    &=\frac{1}{2\pi}\int_{-\infty}^{\infty}e^{-iyt}\frac{e^{-\frac{1}{2}\sigma^2t^t}}{(1+\frac{s}{d} t^2)^n}\text{d}t\\
          &=\frac{1}{2\pi}\int_{-\infty}^{\infty}\frac{e^{-\frac{1}{2}\sigma^2t^2-iyt}}{(1+\frac{s}{d} t^2)^n}\text{d}t,\\ 
\mathbb{P}(|y|<\epsilon) &=\int_{-\epsilon}^{\epsilon}\frac{1}{2\pi}\int_{-\infty}^{\infty}\frac{e^{-\frac{1}{2}\sigma^2t^2-iyt}}{(1+\frac{s}{d} t^2)^n}\text{d}t \text{ d}y .       
\end{align*}
It can be shown that $P_{\epsilon,d}(x_j)$ (4) becomes
\begin{align*}
&P_{\epsilon,d}(x_j) = \sum_{n=0}^{N-1} \int_{-\epsilon-\sqrt{\frac{s}{d}}x_j}^{\epsilon-\sqrt{\frac{s}{d}}x_j}\frac{1}{2\pi}\int_{-\infty}^{\infty}\frac{e^{-\frac{1}{2}\sigma^2t^2-iyt}}{(1+\frac{s}{d}t^2)^n}\text{d}t\text{ d}y\cdot\binom{N-1}{n}\left( \frac{d}{N}\right)^n\left( 1-\frac{d}{N}\right)^{N-1-n}\cdot\frac{d}{sN}.
\end{align*}

\section*{ACKNOWLEDGMENTS}
We thank Ilya Poltorak for useful discussions that led to the development of Noisy-Sudocodes; 
Wenbin Zhu for suggesting that we utilize the characteristic function for the proof of asymptotic independence in Lemma~1;
Junan Zhu for providing us with Matlab code to calculate the MMSE for the matrix channel~\cite{RFG2012,GuoBaronShamai2009}; 
Jin Tan for suggestions that greatly improved our work;
and the reviewers for their careful evaluation of the manuscript. 
\bibliographystyle{IEEEtran}
\bibliography{cites}

\begin{thebibliography}{10}
\providecommand{\url}[1]{#1}
\csname url@samestyle\endcsname
\providecommand{\newblock}{\relax}
\providecommand{\bibinfo}[2]{#2}
\providecommand{\BIBentrySTDinterwordspacing}{\spaceskip=0pt\relax}
\providecommand{\BIBentryALTinterwordstretchfactor}{4}
\providecommand{\BIBentryALTinterwordspacing}{\spaceskip=\fontdimen2\font plus
\BIBentryALTinterwordstretchfactor\fontdimen3\font minus
  \fontdimen4\font\relax}
\providecommand{\BIBforeignlanguage}[2]{{%
\expandafter\ifx\csname l@#1\endcsname\relax
\typeout{** WARNING: IEEEtran.bst: No hyphenation pattern has been}%
\typeout{** loaded for the language `#1'. Using the pattern for}%
\typeout{** the default language instead.}%
\else
\language=\csname l@#1\endcsname
\fi
#2}}
\providecommand{\BIBdecl}{\relax}
\BIBdecl

\bibitem{MaBaronNeedell2013}
Y.~Ma, D.~Baron, and D.~Needell, ``Two-part reconstruction in compressed
  sensing,'' in \emph{Proc. IEEE Global Conf. Signal Inf. Process.}, Austin,
  TX, Dec. 2013.

\bibitem{DonohoCS}
D.~Donoho, ``Compressed sensing,'' \emph{IEEE Trans. Inf. Theory}, vol.~52,
  no.~4, pp. 1289--1306, Apr. 2006.

\bibitem{CandesRUP}
E.~Cand\`{e}s, J.~Romberg, and T.~Tao, ``Robust uncertainty principles: {E}xact
  signal reconstruction from highly incomplete frequency information,''
  \emph{IEEE Trans. Inf. Theory}, vol.~52, no.~2, pp. 489--509, Feb. 2006.

\bibitem{C06:Compressive}
E.~J. Cand\`es, ``Compressive sampling,'' in \emph{Proc. Int. Congress of
  Mathematics}, vol.~3, Madrid, Spain, 2006, pp. 1433--1452.

\bibitem{Indyk2008}
P.~Indyk, ``Explicit constructions for compressed sensing of sparse signals,''
  in \emph{Proc. 19th ACM-SIAM Symp. Discrete Algos.}, Jan. 2008, pp. 30--33.

\bibitem{Iwen2014sparsematrice}
M.~A. Iwen, ``Compressed sensing with sparse binary matrices: Instance optimal
  error guarantees in near-optimal time,'' \emph{Journal of Complexity},
  vol.~30, Feb. 2014.

\bibitem{jafarpour2009}
S.~Jafarpour, W.~Xu, B.~Hassibi, and R.~Calderbank, ``Efficient and robust
  compressed sensing using optimized expander graphs,'' \emph{IEEE Trans. Inf.
  Theory}, vol.~55, no.~9, pp. 4299--4308, Sept. 2009.

\bibitem{Raginsky2011expander}
M.~Raginsky, S.~Jafarpour, Z.~Harmany, R.~Marcia, R.~Willett, and
  R.~Calderbank, ``Performance bounds for expander-based compressed sensing in
  {P}oisson noise,'' \emph{IEEE Trans. Signal Process.}, vol.~59, no.~9, pp.
  4139--4153, Sept. 2011.

\bibitem{Candes05b}
E.~J. Cand\`{e}s, J.~Romberg, and T.~Tao, ``Stable signal recovery from
  incomplete and inaccurate measurements,'' \emph{Comm. Pure Appl. Math.},
  vol.~59, pp. 1207--1223, Mar. 2006.

\bibitem{Cosamp08}
D.~Needell and J.~A. Tropp, ``Co{S}a{MP}: Iterative signal recovery from
  incomplete and inaccurate samples,'' \emph{Appl. Comput. Harm. Anal.},
  vol.~26, no.~3, pp. 301--321, May 2009.

\bibitem{BlumensathDavies2009}
T.~Blumensath and M.~E. Davies, ``Iterative hard thresholding for compressed
  sensing,'' \emph{Appl. Comput. Harm. Anal.}, vol.~27, no.~3, pp. 265--274,
  Nov. 2009.

\bibitem{SudoLDPC}
S.~Sarvotham, D.~Baron, and R.~G. Baraniuk, ``Compressed sensing reconstruction
  via belief propagation,'' Rice University, Houston, TX, Tech. Rep. TREE0601,
  July 2006.

\bibitem{CSBP2010}
D.~Baron, S.~Sarvotham, and R.~G. Baraniuk, ``Bayesian compressive sensing via
  belief propagation,'' \emph{IEEE Trans. Signal Process.}, vol.~58, pp.
  269--280, Jan. 2010.

\bibitem{DMM2009}
D.~L. Donoho, A.~Maleki, and A.~Montanari, ``{Message passing algorithms for
  compressed sensing},'' \emph{Proc. Nat. Acad. Sci.}, vol. 106, no.~45, pp.
  18\,914--18\,919, Nov. 2009.

\bibitem{Bayati2011}
M.~Bayati and A.~Montanari, ``The dynamics of message passing on dense graphs,
  with applications to compressed sensing,'' \emph{IEEE Trans. Inf. Theory},
  vol.~57, no.~2, pp. 764--785, Feb. 2011.

\bibitem{Javanmard2012}
A.~Javanmard and A.~Montanari, ``Subsampling at information theoretically
  optimal rates,'' in \emph{Proc. Int. Symp. Inf. Theory (ISIT)}, July 2012,
  pp. 2431--2435.

\bibitem{Barbier2014}
J.~Barbier, F.~Krzakala, and C.~Schulke, ``Compressed sensing and approximate
  message passing with spatially-coupled {F}ourier and {H}adamard operators,''
  \emph{Arxiv preprint arXiv:1312.1740}, Mar. 2014.

\bibitem{sudo_isit}
S.~Sarvotham, D.~Baron, and R.~G. Baraniuk, ``Sudocodes -- {F}ast measurement
  and reconstruction of sparse signals,'' in \emph{Proc. Int. Symp. Inf. Theory
  (ISIT2006)}, Seattle, WA, July 2006.

\bibitem{Luby2005}
M.~Luby and M.~Mitzenmacher, ``Verification-based decoding for packet-based
  low-density parity-check codes,'' \emph{IEEE Trans. Inf. Theory}, vol.~20,
  no.~1, pp. 120--127, Jan. 2005.

\bibitem{ZhangPfister2012}
F.~Zhang and H.~D. Pfister, ``Veriﬁcation decoding of high-rate {LDPC} codes
  with applications in compressed sensing,'' \emph{IEEE Trans. Inf. Theory},
  vol.~58, no.~8, pp. 5042--5058, Aug. 2012.

\bibitem{Talari2011gbcs}
A.~Talari and N.~Rahnavard, ``{GBCS}: {A} two-step compressive sensing
  reconstruction based on group testing and basis pursuit,'' in \emph{Military
  Comm. Conf.}, Nov. 2011, pp. 157--162.

\bibitem{Boufounos2008}
P.~Boufounos and R.~Baraniuk, ``{1-bit compressive sensing},'' in \emph{Proc.
  2008 Conf. Inf. Sciences Systems}, Mar. 2008, pp. 16--21.

\bibitem{Papoulis91}
A.~Papoulis, \emph{Probability, Random Variables, and Stochastic
  Processes}.\hskip 1em plus 0.5em minus 0.4em\relax McGraw Hill Book Co.,
  1991.

\bibitem{RanganGAMP2010}
S.~Rangan, ``Generalized approximate message passing for estimation with random
  linear mixing,'' \emph{Arxiv preprint arXiv:1010.5141}, Oct. 2010.

\bibitem{Montanari2012}
A.~Montanari, ``Graphical models concepts in compressed sensing,''
  \emph{Compressed Sensing: Theory and Applications}, pp. 394--438, 2012.

\bibitem{Krzakala2012probabilistic}
F.~Krzakala, M.~M{\'e}zard, F.~Sausset, Y.~Sun, and L.~Zdeborov{\'a},
  ``Probabilistic reconstruction in compressed sensing: {A}lgorithms, phase
  diagrams, and threshold achieving matrices,'' \emph{J. Stat. Mech. - Theory
  E.}, vol. 2012, no.~08, p. P08009, Aug. 2012.

\bibitem{ZhuBaronCISS2013}
J.~Zhu and D.~Baron, ``Performance regions in compressed sensing from noisy
  measurements,'' in \emph{Proc. 2013 Conf. Inf. Sciences Systems}, Baltimore,
  MD, Mar. 2013, pp. 1--6.

\bibitem{RFG2012}
S.~Rangan, A.~K. Fletcher, and V.~K. Goyal, ``Asymptotic analysis of {MAP}
  estimation via the replica method and applications to compressed sensing,''
  \emph{IEEE Trans. Inf. Theory}, vol.~58, no.~3, pp. 1902--1923, Mar. 2012.

\bibitem{GuoBaronShamai2009}
D.~Guo, D.~Baron, and S.~Shamai, ``A single-letter characterization of optimal
  noisy compressed sensing,'' in \emph{Proc. 47th Allerton Conf. Commun.,
  Control, Comput.}, Sept. 2009, pp. 52--59.

\bibitem{Jacques2011robust}
L.~Jacques, J.~N. Laska, P.~T. Boufounos, and R.~G. Baraniuk, ``Robust 1-bit
  compressive sensing via binary stable embeddings of sparse vectors,''
  \emph{IEEE Trans. Inf. Theory}, vol.~59, Apr. 2013.

\bibitem{PlanVershynin2012}
Y.~Plan and R.~Vershynin, ``One-bit compressed sensing by linear programming,''
  \emph{Comm. Pure Appl. Math.}, vol.~66, pp. 1275--1297, Aug. 2013.

\bibitem{Laska2011trust}
J.~N. Laska, Z.~Wen, W.~Yin, and R.~G. Baraniuk, ``Trust, but verify: Fast and
  accurate signal recovery from 1-bit compressive measurements,'' \emph{IEEE
  Trans. Signal Process.}, vol.~59, no.~11, pp. 5289--5301, Nov. 2011.

\bibitem{Yan2012robust1bit}
M.~Yan, Y.~Yang, and S.~Osher, ``Robust 1-bit compressive sensing using
  adaptive outlier pursuit,'' \emph{IEEE Trans. Signal Process.}, vol.~60,
  no.~7, pp. 3868--3875, July 2012.

\bibitem{Yang2013variational}
Z.~Yang, L.~Xie, and C.~Zhang, ``Variational {B}ayesian algorithm for quantized
  compressed sensing,'' \emph{IEEE Trans. Signal Process.}, vol.~61, no.~11,
  pp. 2815--2824, June 2013.

\bibitem{Applebaum2005}
D.~Applebaum, B.~Bhat, J.~Kustermans, and J.~Lindsay, \emph{Quantum Independent
  Increment Processes I: From Classical Probability to Quantum Stochastic
  Calculus (Lecture Notes in Mathematics)}.\hskip 1em plus 0.5em minus
  0.4em\relax New York, NY, USA: Springer, Feb. 2005.

\bibitem{Klenke2013}
A.~Klenke, \emph{Probability Theory: A Comprehensive Course}.\hskip 1em plus
  0.5em minus 0.4em\relax New York, NY, USA: Springer, Sept. 2013.

\end{thebibliography}
\end{document}